# Linear accelerators

*Maurizio Vretenar*
CERN, Geneva, Switzerland

**Abstract**
Radio-frequency linear accelerators are used as injectors for synchrotrons and as stand-alone accelerators for the production of intense particle beams, thanks to their ability to accelerate high beam currents at high repetition rates. This lecture introduces their main features, reviewing the different types of accelerating structures used in linacs and presenting the main characteristics of linac beam dynamics. Building on these bases, the architecture of modern proton linear accelerators is presented with a particular emphasis on high-energy and high-beam-power applications.

## 1 Introduction, general features

A *linear accelerator* (linac) is a device where charged particles acquire energy moving on a linear path; the characteristic feature of a linac is that the particles pass only once through each of its accelerating structures.[1] In the following, we limit our analysis to radio-frequency (RF) linacs where the acceleration is provided by time-varying electric fields, leaving out of our treatment *electrostatic linacs* that are usually limited to energies outside of the scope of this lecture. Moreover, we concentrate on linacs for high-beam-power applications and therefore exclude the large and important domain of *electron linacs*.

A linac will take the continuous particle beam coming out of an ion source, bunch it a given RF frequency and then accelerate it up to the required final energy. In general, linacs are *pulsed* accelerators: the beam is generated by the source and then delivered to the users in pulses of a given length $\tau$ (between few microseconds and few milliseconds) at a given repetition frequency $f$ (usually between 1 Hz and 100 Hz). The product of pulse length and repetition frequency is the *duty cycle* (or *beam duty cycle*, to distinguish it from the *RF duty cycle* which is always higher). A linac can as well operate continuously, producing a constant stream of particles: in this case the duty cycle is 100%, and we call it a continuous wave (CW) linac.

Together with the *kinetic energy E* of the particles coming out of the linac, its most important parameter is the beam current $I$, defined as the *average current during the beam pulse*. The current $I$ is different from the average current out of the linac, which is $I$ times the duty cycle; it can also be different from the *bunch current*, the average current during a RF period, in the particular case where not all bunches are populated by particles.

The *beam power P* defines the electrical power (measured in Watts) transferred to the particle beam during the acceleration process. It represents the sum of the electrical power absorbed by the beam in the different accelerating cavities that constitute the accelerator. In each cavity, the current $I$ crosses a voltage $V$; the RF power going to the beam is $P = V \times I$, considering the beam current as constant during the pulse at the value $I$ and taking for $V$ the voltage actually seen by the beam, corrected with the transit time factor of the particles in the cavity gap.[2]

[1] It should be mentioned that linacs do not need to be straight; some heavy ion linacs for example incorporate 90° bends to reduce the footprint of the machine. In a similar way, some electron linac designs can include curved sections and even pass the beam two or more times through the same accelerating structures; these are the "*recirculating linacs*".

Under these assumptions, the overall voltage $V_{tot}$, the sum of the effective voltage seen by the beam in all the RF cavities, is (numerically) equal to $E$, the total beam energy expressed in electronvolts, thus the total beam power produced by a linac is

$$P\,[\mathrm{W}] = E\,[\mathrm{eV}] \times I\,[\mathrm{A}] \times \text{duty cycle}$$

The beam power is important because on the one hand it represents the amount of power that the RF system has to deliver to the beam, and on the other hand it corresponds to a simple "figure of merit" for linacs dedicated to the production of secondary particles (neutrons, pions, etc.). Above a specific energy threshold, the number of secondary particles produced by the target is directly proportional to the primary beam power (i.e. to the product of beam intensity and energy).

For the production of high-power beams, linear accelerators have many advantages as compared with other types of accelerators. The repetition frequency of a linac is not limited by the rise time of the magnets as is the case for synchrotrons. It can therefore reach very high values, going up to CW operation; for example, several high-power linacs adopt as basic repetition frequency that of the mains supply, 50 Hz in Europe and 60 Hz in the US. Moreover, the fact that the beam passes only once in each section of a linac limits the effect on the beam of magnetic field errors: beam resonances that constitute the main limitation of synchrotrons in achieving high beam currents have only a minor impact on linear accelerators. Cyclotrons are often in competition with linacs for the production of low-energy CW proton and ion beams; however, their advantages in terms of compactness and cost vanish for energies above the relativistic limit where special designs are required. Linear accelerators can easily reach energies in the GeV range, although their relatively high cost per MeV of acceleration suggests the transition to a synchrotron for very high beam energies when the synchrotron repetition frequency is compatible with the beam parameter of the project. It should be observed that the optimum transition energy between a linac and a rapid cycling synchrotrons (RCS), the special type of synchrotron required for high beam power, is very difficult to define and depends on the specific beam parameters as well as on the experience of the particular laboratory.

The role of linacs with respect to cyclotrons and synchrotrons can be easily visualized considering the proton velocity curve of Fig. 1, where $\beta$ = v/c is plotted as a function of kinetic energy.

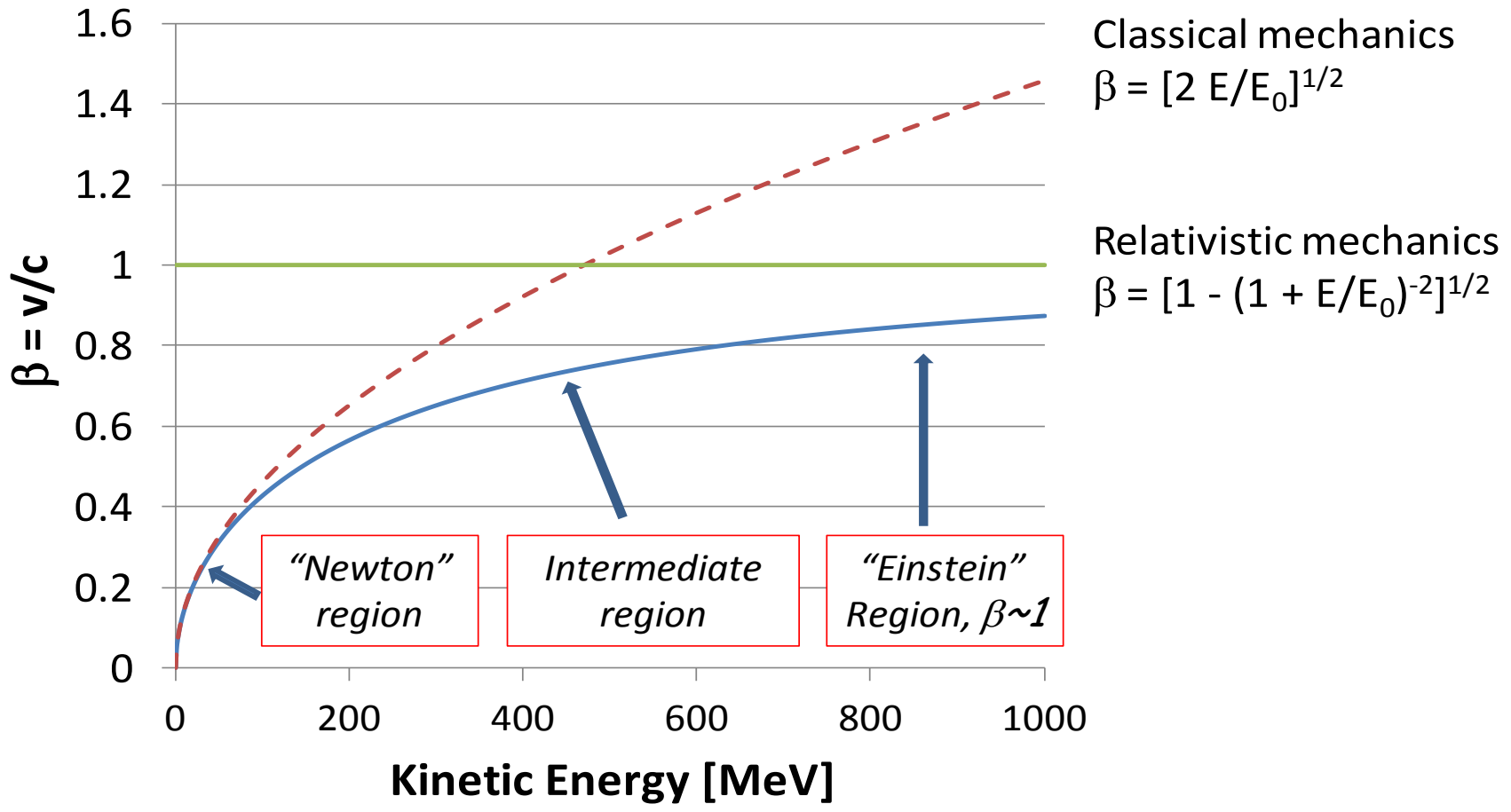

**Fig. 1:** Relativistic velocity as a function of kinetic energy (protons)

[2] Actually, the calculation is a bit more complicated: the voltage is sinusoidal, and only the Fourier component of the beam current at the RF frequency absorbs power following the usual formula $P = \frac{1}{2}V \times i_{RF}$; but for bunches that are short with respect to the RF period, the amplitude of the main harmonic component is $i_{RF} = 2I$, giving the well-known relation $P = V \times I$.

At the beginning of acceleration the particle velocity follows the classical square root relation, but already at proton energies of a few tens of megaelectronvolts the velocity starts to be lower than that predicted by classical mechanics; the beam is now in the domain of relativistic mechanics. From the gigaelectronvolt range, the velocity increases very slowly and starts approaching asymptotically the speed of light: the energy that our accelerator delivers to the beam goes into increasing the mass of the particles instead of increasing their velocity.

As we will see in the following, a linear accelerator is made of cells of variable length. The main feature of linacs is that by adapting the cell lengths to the increase in particle velocity one can synchronize the acceleration with the selected RF frequency. The procedure consists of defining the cell lengths in such a way that the time needed by the beam to cross a cell remains constant; the result is that the cell length will progressively increase with the energy. A linear accelerator can adapt to any beam velocity: it can cover at the same time the "Newton" range of energies, where the velocity increases with the energy, as well as the "Einstein" range, where the velocity remains approximately constant. A synchrotron instead is a typical "Einstein" machine, designed for a fixed revolution frequency and for a beam of almost constant velocity. Actually, all synchrotrons allow for a modulation of the RF frequency which permits the acceleration of beams of increasing velocity; however, the modulation range is usually small and large modulations require special and expensive RF cavities. For this reason, a linac is always used as an injector to a synchrotron, covering the energy range where the velocity variation is large while the synchrotron takes over when the variation becomes smaller. The "intermediate" region, between the Newton and Einstein regimes, can be covered by both linacs and synchrotrons: if high beam intensity and/or high beam quality are required, the high repetition frequency and the absence of resonances give an advantage to linacs, while if the objective consists in reaching a high-energy synchrotrons have an advantage because their cost per unit of acceleration is lower than that of linacs. Cyclotrons instead are special accelerators where the beam remains synchronous with the RF only in the "Newton" range. Their compactness and reduced cost give them an advantage with respect to linacs as a stand-alone machine for producing non-relativistic CW beams; to achieve higher energies cyclotrons require special technologies that largely reduce their attractiveness.

As we have mentioned, a linear accelerator is made of a sequence of accelerating gaps as in Fig. 2, where for convenience each gap is associated with an individual RF cavity.

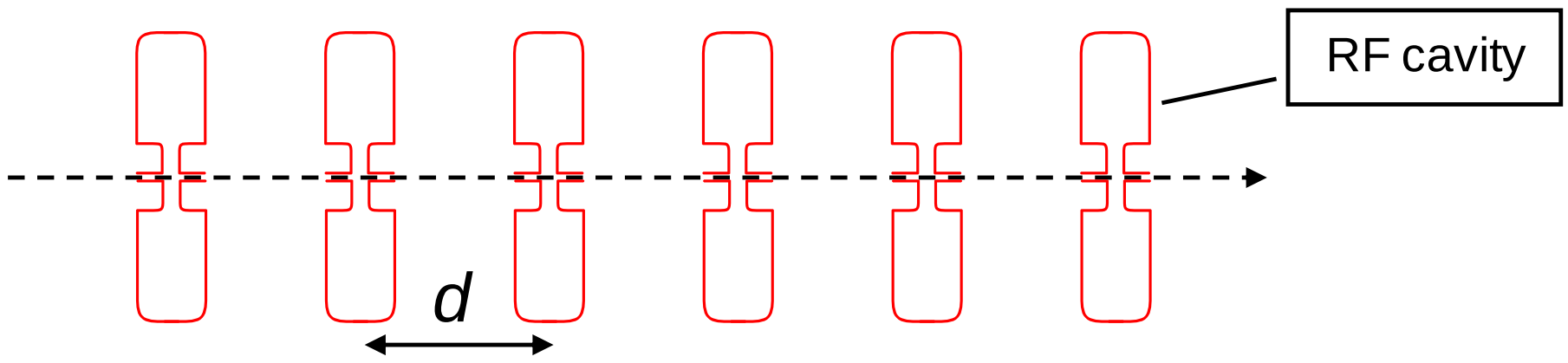

**Fig. 2:** Sequence of accelerating gaps/cavities

If we want to accelerate a beam of increasing velocity we can immediately observe that: (a) *the beam must be already bunched at the RF frequency when it enters the sequence*; and (b) *for a beam of increasing velocity, the distance between the cavities and their relative RF phase must be correlated*. As usual, the electric field on the gap of cavity *i* can be written as

$$E_i = E_{0i} \cos(\omega t + \varphi_i)$$

with $\varphi_i$ the phase of the *i*th cavity with respect to a "reference" RF phase. To maximize acceleration, the beam has to cross the gap of each cavity at a phase $\varphi_i$ on or very close to the crest of the wave ($\varphi_i = 0$); moreover, it must have a short length in time and in phase, i.e. it must be "bunched". During the time that the particles need to go from one cavity to the next the phase has changed by an amount $\Delta\varphi = \omega\tau$ with $\tau$ the time to cross the distance *d*; for a particle of relativistic velocity $\beta = v/c$ the change in phase will be

$$\Delta\Phi = \omega\tau = \omega\frac{d}{\beta c} = 2\pi\frac{d}{\beta\lambda}$$

Here *d* is the distance between gaps and $\lambda$ the RF wavelength. This means that

$$\frac{\Delta\Phi}{d} = \frac{2\pi}{\beta\lambda}$$

or, for the acceleration to take place, the *distance* and the *phase difference* between two gaps in the sequence must be correlated, their ratio being proportional to $\beta\lambda$. At every gap crossing the particle will gain some energy and its velocity will increase: in a non-relativistic regime this means that either the relative phase $\Delta\Phi$ or the distance *d* has to change during acceleration. In other terms, in a linear accelerator we need either to progressively increase the distance between cavities or to progressively decrease their RF phase (relative to a common reference) to keep synchronicity between the particle beam and the accelerating wave.

This requirement corresponds to two well-defined types of linear accelerators:

1. “Single-cavity” linacs (Fig. 3, top), where the *distance* between cavities is fixed, and the phase of each cavity is individually adjusted to take into account the increase in beam velocity; each cavity has to be connected to an individual RF amplifier. This scheme has the advantage of maximum flexibility, being able to accelerate different ions and/or charge states at different energies by entering a different set of phases for each ion, but has the drawback of a high cost.

2. “Coupled-cell cavity” linacs (Fig. 3, bottom), where the *phase* at each cavity/gap is fixed, and the distance changes accordingly to the beam velocity. When the RF phase is the same for each gap, several gaps can be coupled into a common RF structure and connected to a single RF power source, reducing significantly the cost of the RF system. The single gaps and resonators have to be coupled together, however, to allow fixing their relative phase (this will be the subject of the next section). This scheme is cost effective but not at all flexible: the physical distance between gaps is defined for a given energy increase, i.e. for a given particle, energy range and acceleration gradient.

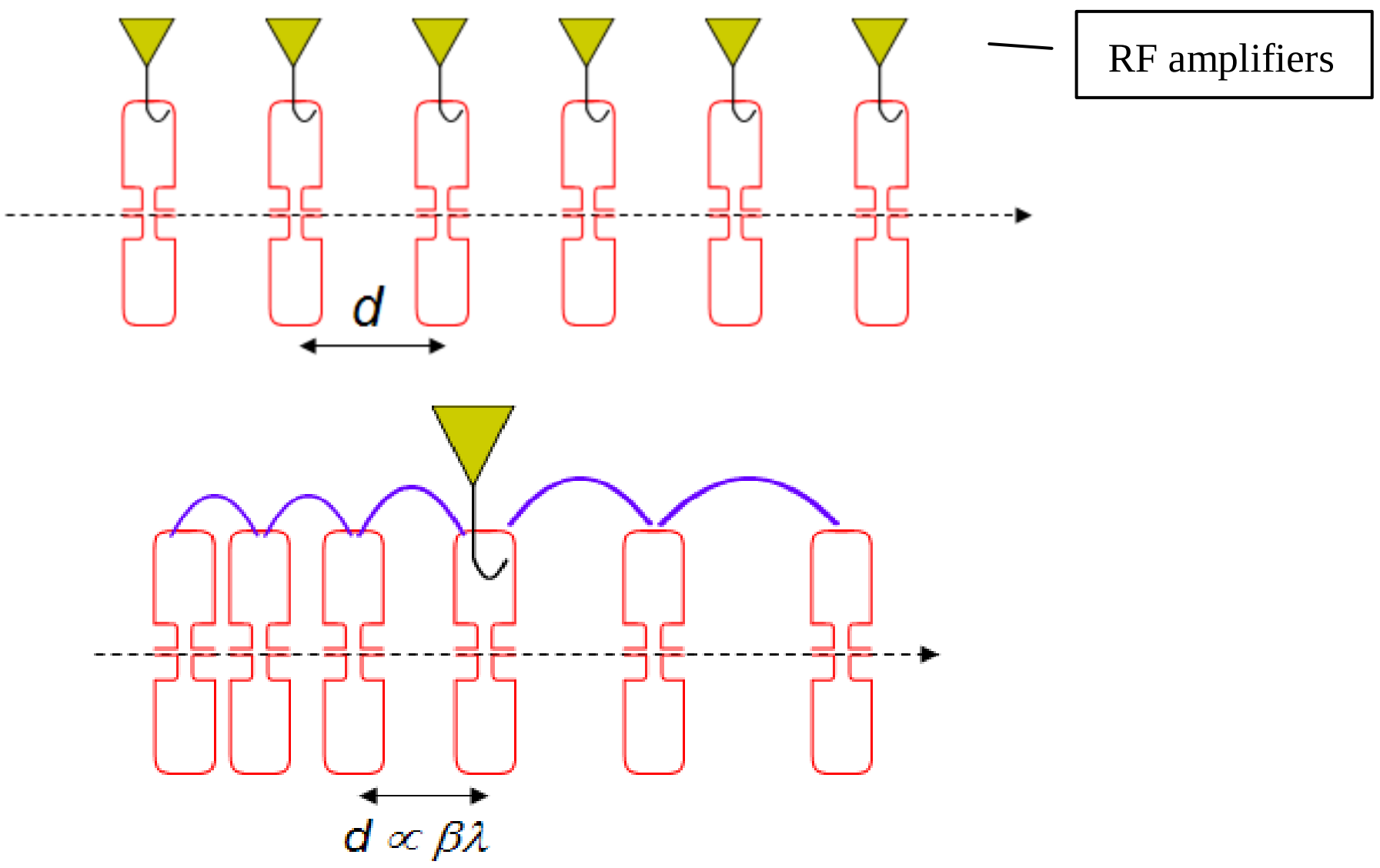

**Fig. 3:** Single-cavity and coupled-cell-cavity linacs

The reality of linac structures is not so antithetic, however, and most of the commonly used linacs tend to compromise between these two extreme configurations. This is particularly true at high energies: if the increase in $\beta$ between two gaps is small, the phase error in the case where the gaps are equally spaced will be small and under certain circumstances acceptable. This is the case of multicell superconducting cavities such as that in Fig. 4, commonly used in the high-energy section of linacs. The structure of Fig. 4 is made of four gaps coupled together in a single resonator. At a given time, the longitudinal electric field on the axis of the cavity will have the profile shown in the figure and indicated by the arrows in the gaps: this corresponds to a constant phase difference $\Delta\Phi$ = 180° between adjacent gaps. For a particle to be accelerated, the relation between phase and gap distance must hold, with $\Delta\Phi = \pi$; solving for $d$, we find that the distance between gaps must be equal to $\beta\lambda/2$. This condition is respected only for one particular $\beta$, however; inside the cavity the energy and the beta are increasing, leading to a difference in the phase of the RF seen by the beam in the different cells that is called "phase slippage". For this structure to accelerate effectively, the phase slippage must be small as compared with the synchronous phase $\varphi_i$, i.e. the increase in $\beta$ within the cavity must be small.

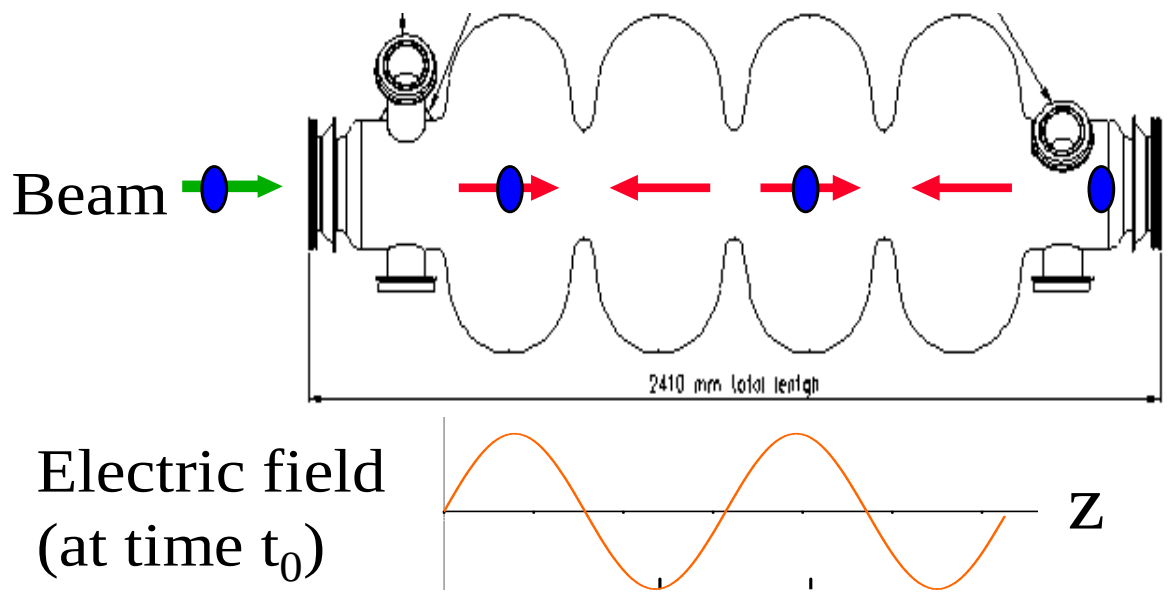

**Fig. 4:** Multicell superconducting cavity

Before going into the study of the different types of accelerating structures, it is important to consider that a linac is much more complex than a simple sequence of RF gaps. First of all, the sequence of gaps, usually grouped inside a RF cavity, has to be preceded by an ion source and by a bunching system[3]. Then, the gaps have to be spaced by some focusing elements, usually quadrupoles, required to keep together the particles that constitute the bunch. The RF cavity needs to be fed by a RF amplifier inserted into a feedback loop and the system requires some ancillaries to work: a vacuum system, a magnet powering system and a water cooling system to evacuate the excess power provided by the RF system. A basic block diagram of a linac system is presented in Fig. 5.

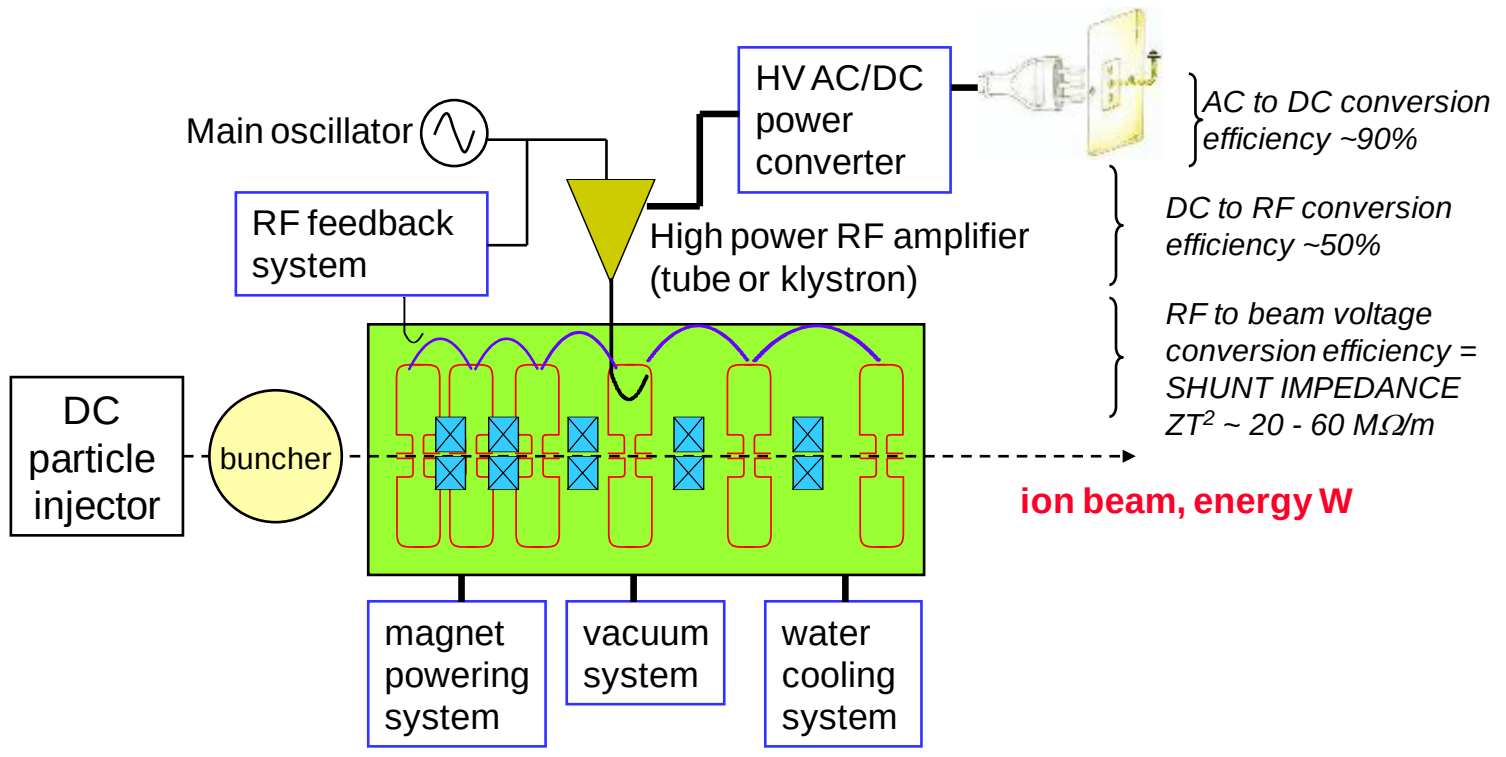

**Fig. 5:** Scheme of a linac accelerating system

[3] The different bunching mechanisms are described in the lecture on RFQs in these proceedings.

The representation of Fig. 5 visualizes the fact that a linac RF system transforms electrical energy taken from the grid into energy transferred to a beam of particles. The corresponding power is the beam power introduced at the beginning of this lecture. The efficiency of this transformation is usually quite low and a large fraction of the input energy is dissipated into heat released into the surrounding environment. More precisely, the energy transformation takes place in three different steps, each one characterized by a different technology and a different efficiency:

(i) The transformation of the AC power from the grid (alternate, low voltage, high current) in DC power (continuous, high voltage, low current) takes place in a power converter. Pulsed power converters are usually called "modulators"; their efficiency is usually high, of the order of 80 % to 90 %.

(ii) The following transformation of the DC power into RF power (high frequency, high voltage, low current) takes place in a RF active element: RF tube, klystron, transistor, etc.; the efficiency is the RF conversion efficiency that depends on the specific device and on its class of operation. Typical RF efficiencies are in the 50 % to 60 % range.

(iii) The final transformation of the RF power into power stored into the particle beam takes place in the gap of an accelerating cavity; the efficiency is proportional to the shunt impedance of the cavity, which represents the efficiency of the gap in converting RF power into voltage available for a beam crossing the cavity at a given velocity (see Section 6).

## 2 Accelerating structures for linacs

Coupled-cell cavities are the most widely used accelerating structures for linacs. To couple the elements of a chain of single-gap resonators (that from now on we will often refer to as the "cells" of our system) we need to allow some energy to flow from one cell to the next, via an aperture that permits leaking of some field (electric or magnetic) into the adjacent resonator. There will be two different types of coupling, depending on whether the opening connects regions of high magnetic field ("magnetic coupling") or regions of high electric field ("electric coupling"). The simplest magnetic coupling is obtained by opening a slot on the outer contour of the cell, whereas an electric coupling can be obtained by enlarging the beam hole until some electric field lines couple from one cell to the next. Once the cells are coupled, to find the conditions for acceleration we have to calculate the relative RF phase of the individual cells.

The simplest way to analyse the behaviour of a chain of coupled oscillators is to consider their equivalent circuits (Fig. 6) [1].

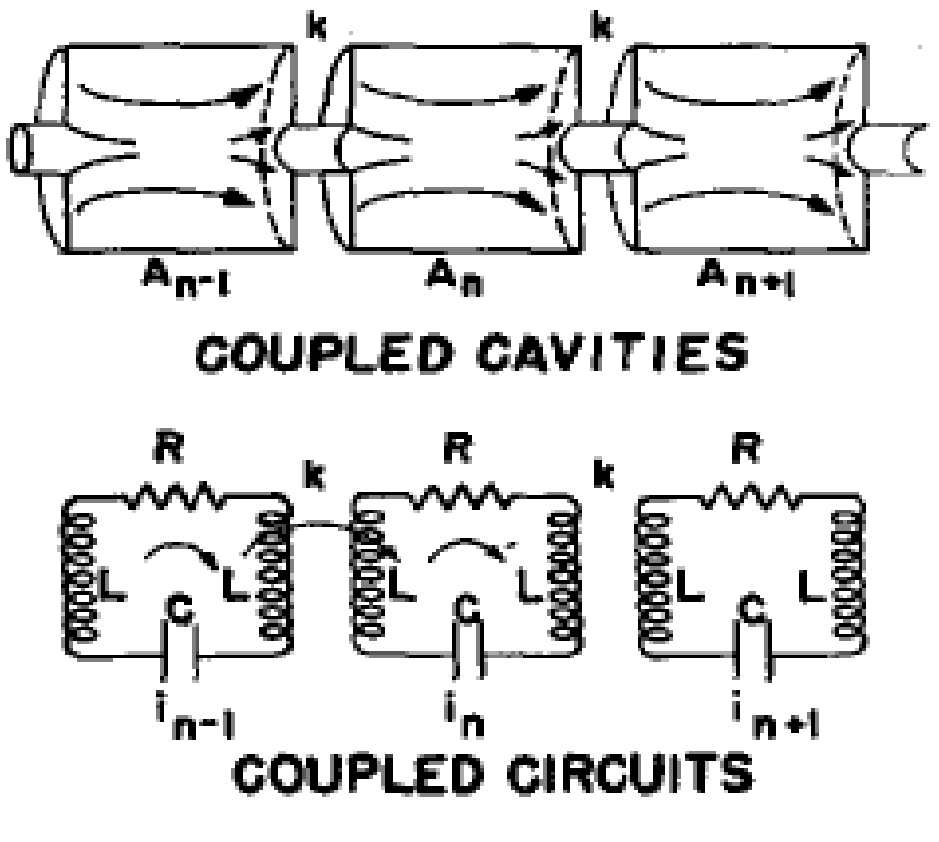

**Fig. 6:** From coupled cavities to coupled resonant electrical circuits (from Ref. [1])

Each coupled cavity can be represented by a standard *RLC* resonant circuit; for convenience, in Fig. 6 the inductance of each circuit is split into two separated inductances *L*. The advantage of this representation is that we can describe the (magnetic) coupling between adjacent cells as a mutual inductance *M* between two inductances *L*, which is related to a coupling factor *k* by the usual relation $M = kL$. For the series resonant circuits of Fig. 6, the behaviour of each cell is described by its circulating current $I_i$. The equation for the *i*th circuit can be written taking equal to zero the sum of the voltages across the different elements of the circuit (Kirchhoff's law), considering for simplicity a lossless system with $R = 0$:

$$I_i\left(2j\omega L + \frac{1}{j\omega C}\right) + j\omega kL(I_{i-1} + I_{i+1}) = 0$$

Dividing both terms of this equation by $2j\omega L$, it can be written as

$$X_i\left(1 - \frac{\omega_0^2}{\omega^2}\right) + \frac{k}{2}(X_{i-1} + X_{i+1}) = 0$$

This equation relates general excitation terms of the form $X_i = I_i / 2j\omega L$, proportional to the square root of the energy stored in the cell *i*, with the coupling factor *k* and with a standard resonance term $(1-\omega_0^2/\omega^2)$. We consider that all cells are identical, i.e. that they have the same resonance frequency $\omega_0^2 = 1/2LC$. If our system is composed of $N + 1$ cells, assuming $i = 0,1,...,N$ we can write a system of $N + 1$ equations with $N + 1$ unknowns $X_i$ represented by the following matrix equation:

$$\begin{bmatrix} 1-\frac{\omega_0^2}{\omega^2} & \frac{k}{2} & 0 & & 0 \\ \frac{k}{2} & 1-\frac{\omega_0^2}{\omega^2} & \frac{k}{2} & \cdots & \cdot \\ & & & & 0 \\ & & & & \cdot \\ 0 & \frac{k}{2} & 1-\frac{\omega_0^2}{\omega^2} & & 0 \\ & \vdots & & \ddots & \vdots \\ 0 & 0 & 0 & \cdots & 1-\frac{\omega_0^2}{\omega^2} \end{bmatrix} \begin{vmatrix} X_0 \\ X_1 \\ \cdots \\ X_N \end{vmatrix} = 0$$

or

$$MX = 0$$

The matrix *M* has elements different from zero only on three diagonals, the main one with the resonance terms and the two adjacent ones with the coupling terms. It is perfectly symmetric because we have introduced an additional simplification, closing at both ends our chain of resonators with "half cells", presenting a coupling *k* only on one side and with half the inductance and twice the capacitance of a standard cell (but the same resonant frequency). This corresponds to a physical case where the end resonators are terminated by a conducting wall passing at the centre of the gap, i.e. they are exactly one half of a standard cell. The advantage of this approach is that the matrix and the relative solutions are symmetric and lead to a simple analytical result. In the real case, the chain of resonators is terminated with full cells that need to be tuned to a slightly different frequency to symmetrise the system.

The above matrix equation represents a standard eigenvalue problem, which has solutions only for those $\omega$ giving

$$\det M = 0$$

The eigenvalue equation $\det M = 0$ is an equation of $(N + 1)$th order in $\omega$. Its $N + 1$ solutions $\omega_q$ are the eigenvalues of the problem, which are the resonance modes of the coupled system. Whereas the individual resonators can oscillate only at the frequency $\omega_0$, the coupled system will have $N + 1$ frequencies $\omega_q$, the "modes", with $q = 0,1,...,N$ the index of the mode. To each mode corresponds a solution in the form of a set of $[X_i]_q$, which is the corresponding eigenvector. It is important to observe that the number of modes is always equal to the number of cells in the system.

For the matrix $M$, we can find an analytical expression for the eigenvalues (mode frequencies):

$$\omega_q^2 = \frac{\omega_0^2}{1 + k \cos \frac{\pi q}{N}} \quad q = 0, \ldots, N \tag{1}$$

or, for $k << 1$, which is the operating condition for most of coupled structures commonly used in linacs where it is usually $k \sim 1$ % to 5 %:

$$\omega_q \approx \omega_0 (1 - \tfrac{1}{2} k \cos \tfrac{\pi q}{N}) \qquad q = 0, \ldots, N$$

The corresponding eigenvectors (modes) are

$$X_i^{(q)} = (const) \cos \frac{\pi q i}{N} e^{j\omega_q t} \qquad q = 0, \ldots, N \tag{2}$$

The expression (1) is particularly interesting, because it indicates that each mode $q$ is identified by a “phase”:

$$\Phi_q = \frac{\pi q}{N}$$

The first mode, $q = 0$, has $\Phi = 0$ and frequency $\omega_{q=0} = \frac{\omega_0}{\sqrt{1+k}}$. The last mode, $q = N$, will have $\Phi = \pi$ and frequency $\omega_{q=N} = \frac{\omega_0}{\sqrt{1-k}}$. If we identify each mode by the value of $\Phi_q$ the first will be the “0” mode and the last the “π” mode. All other modes will have frequencies between the 0 and π mode frequencies.

For $k << 1$ the difference between π and 0 mode frequencies is

$$\Delta\omega = \omega_{q=N} - \omega_{q=0} = \omega_0 (\frac{1}{\sqrt{1-k}} - \frac{1}{\sqrt{1+k}}) \approx \omega_0 k$$

i.e. the “bandwidth” of the coupled system is equal to the cell frequency times the coupling factor $k$.

Plotting the frequencies given by Eq. (1) as function of the phase Φ, we obtain curves such as that in Fig. 7, which corresponds to the case of five cells and five modes. This is a typical “dispersion curve”, relating the frequencies of our system with their propagation constant. The permitted frequencies lie on a cosine-like curve, where the modes are represented by points equally spaced in phase. The more cells in the system, the more modes we will have on the curve, until the limit of the continuous: for an infinite number of cells, all of the modes on the curve are allowed.

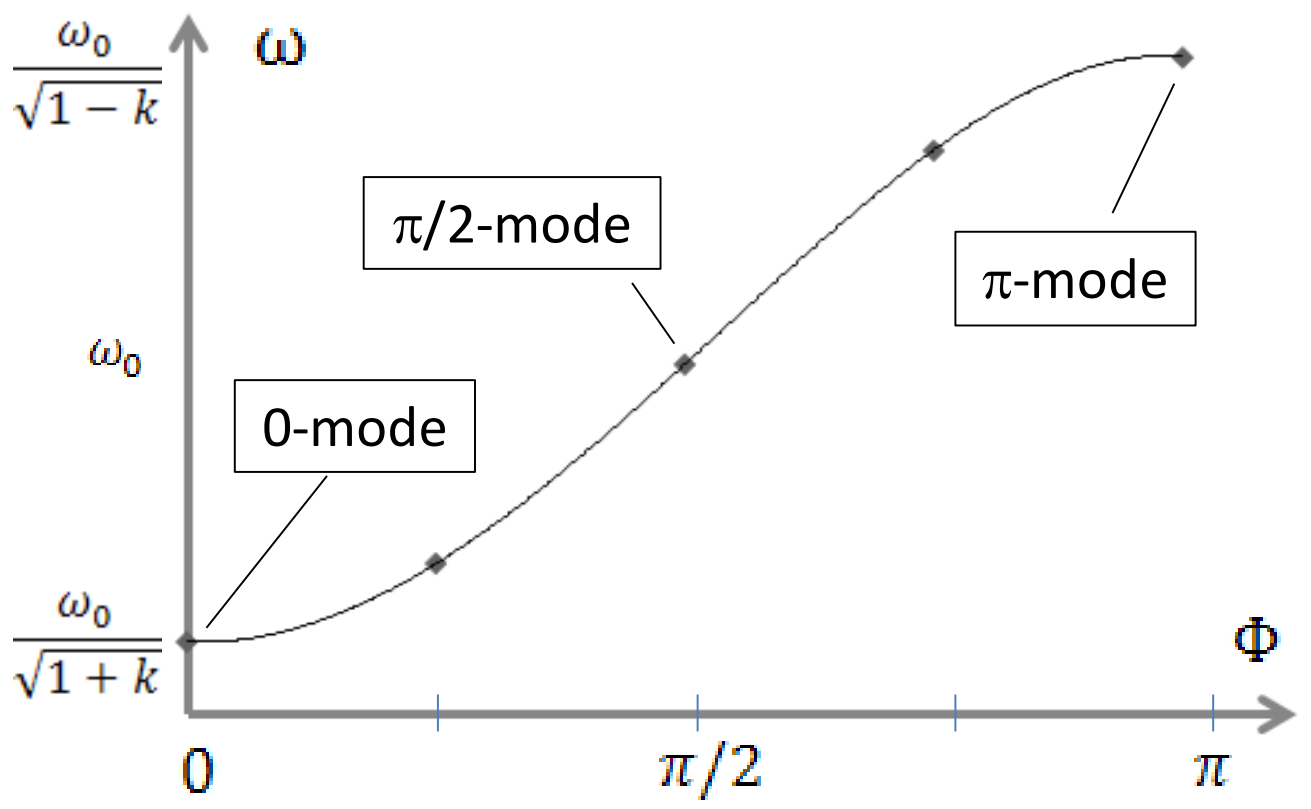

**Fig. 7:** Dispersion relation for a 5-cell coupled resonator chain

The field distribution in the cells is defined by the expression (2). For a given mode $q$, the fields will oscillate in each cell at the frequency $\omega_q$, and *the amplitude of the oscillation will depend on the position of the cell in the chain*. The distribution of maximum field amplitudes along the chain follows a cosine-like function with argument ($\Phi_q i$), i.e. the product of the phase $\Phi_q$ times the cell number $i$. It is now clear that $\Phi_q$ represents the *phase difference between adjacent cells* in the coupled system. We can now draw the field distribution between the cells in the chain for the main modes, for example for a seven-cell system with $N$ = 6 (Fig. 8).

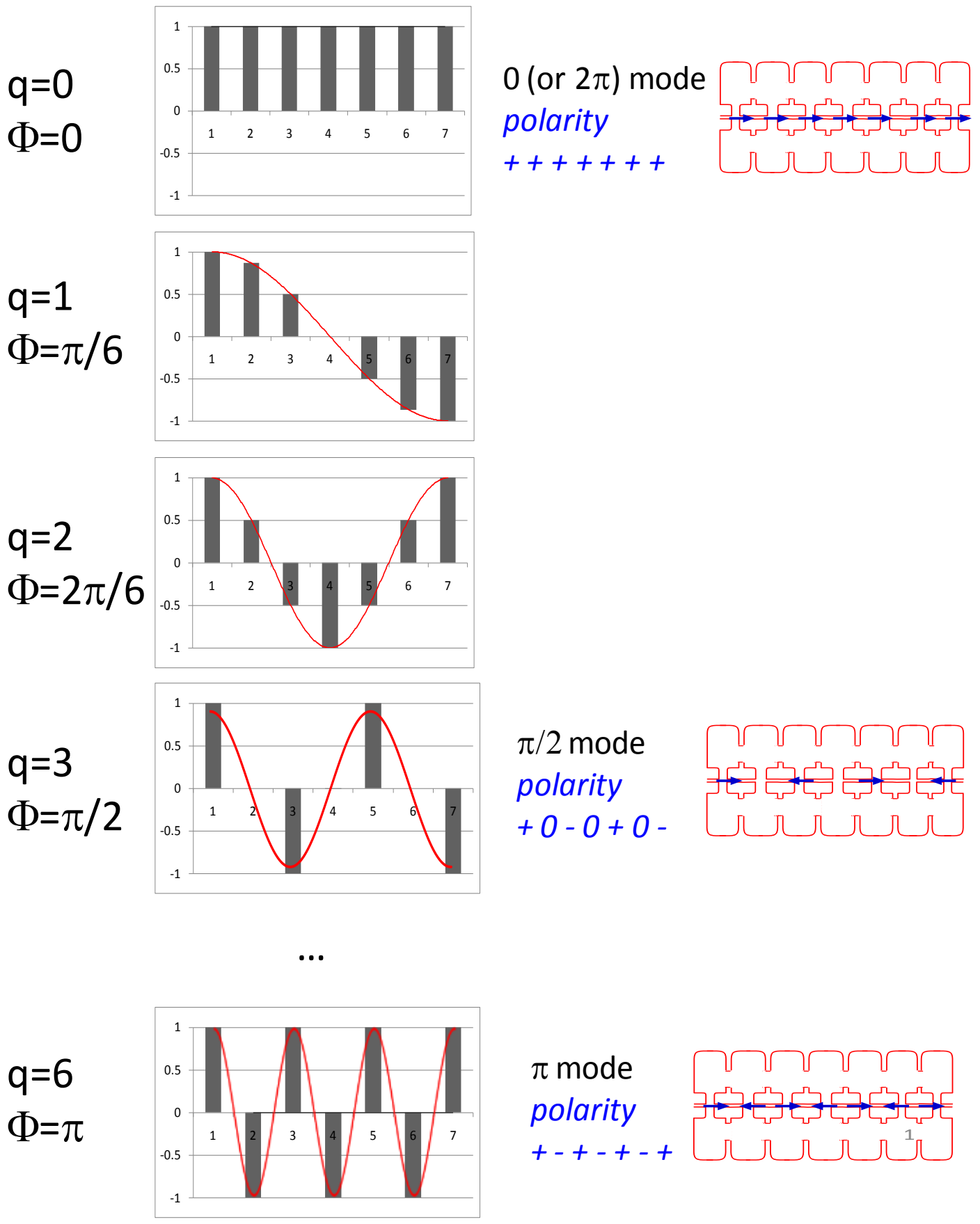

**Fig. 8:** Distribution of the fields in the cells of a seven-cell system and polarity of the electric field in the gaps for the modes used for particle acceleration

We must observe at this point that these are "standing-wave" modes: the plots of Fig. 8 show the field distribution at time $t$ = 0. The fields are oscillating with angular frequency $\omega$, and after $\omega t = \pi/2$ the field that we are plotting will be zero everywhere, whereas after $\omega t = \pi$ it will be maximum again, but with reversed polarity. The modes are identical to those of a vibrating string with the two ends fixed (in our case, defined by the boundary conditions).

To understand which of these modes can be used for the acceleration of particles and under what conditions, we can write the electric field for the mode $q$ at the centre of gap $i$ using expression (2) and then apply a simple trigonometric transformation:

$$E_i^{(q)} = E_0 \cos \Phi_q i \, \cos \omega_q t = \frac{E_0}{2}\left[\cos(\omega_q t - \Phi_q i) + \cos(\omega_q t + \Phi_q i)\right]$$

The electric field is the sum of two cosine functions. The first is the same as that we have introduced at the beginning of the lecture: for maximum acceleration, its argument must be 0 (or, more precisely, $2n\pi$) for a particle going from one cell to the next in the time $\tau$. This gives $\omega_q \tau = \Phi_q$, leading to the synchronism condition:

$$d = \frac{\beta\lambda}{2}\frac{\Phi_q}{\pi}$$

This gives us the well-known result that the distance between the cells must be related to the beam velocity. In particular, for the 0 and π modes we obtain

$$d = \beta\lambda \qquad (0 - mode, \Phi_{\mathrm{q}} = 2\pi)$$

$$d = \frac{\beta\lambda}{2} \qquad (\pi - mode, \Phi_{\mathrm{q}} = \pi)$$

The second cosine function instead tells us which modes can be used for acceleration: for $\omega_q\tau = \Phi_q$ it becomes equal to $\cos 2\Phi_q$ which is 1 only for $\Phi_q = 0, \pi, 2\pi, \ldots$ . The conclusion is that only the modes 0 and π can be used for efficient particle acceleration. An exception is the π/2 mode, which has $\cos 2\Phi_q = 0$. This can still be used for acceleration (with $d = \beta\lambda/4$), but the acceleration is not very efficient, the field being present only in half of the cells. As we will see in the following, however, the π/2 mode presents the advantage of higher stability against deviation in the individual cell frequencies that justifies its use for some specific accelerating structures.

## 3 Zero-mode structures: the drift tube linac

The first and most important structure operating in the 0-mode is the drift tube linac (DTL), also called an Alvarez linac after the name of its inventor. It can be considered as a chain of coupled cells where the wall between cells has been completely removed to increase the coupling (Fig. 9). A high coupling offers the advantage of a large bandwidth, with sufficient spacing between the modes to avoid dangerous instabilities even when the chain is made of a large number of cells. Moreover, in the particular case of a structure operating in the 0-mode removing the cell-to-cell walls does not influence the power loss in the structure, because the RF currents flow only on the external tank and on the tubes. We must, however, keep some tubes on the axis, called "drift tubes", which hide the particles during the half RF period when the electric field on axis is decelerating. If the diameter of the drift tubes is sufficiently large, they can house focusing quadrupoles, which at low energy are required to keep the beam transversally focused. The drift tubes are suspended to the outer tank by means of supports called stems. The basic structure of a DTL is shown in Fig. 10.

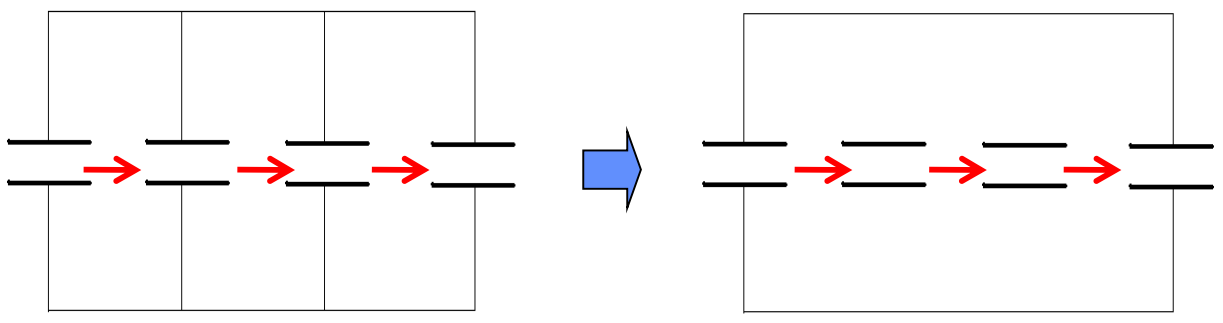

**Fig. 9:** From a chain or resonators operating in 0-mode to the DTL. The arrows indicate the direction of the electric field on axis.

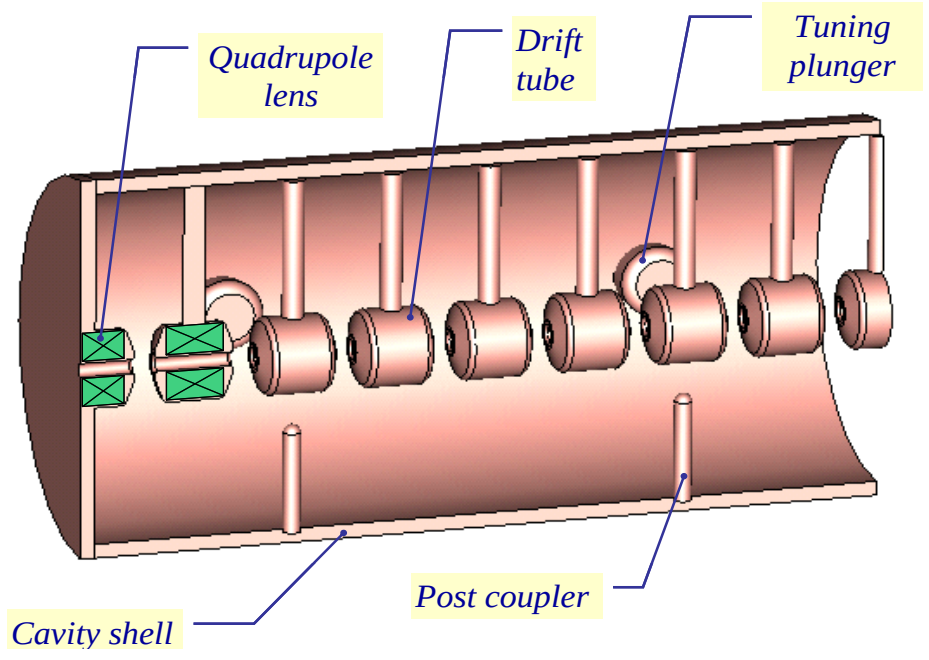

**Fig. 10:** The DTL structure

The DTL can be represented by a coupled circuit model similar to that in Fig. 6. Owing to the absence of the cell walls, however, the coupling mechanism is more complex, resulting in a strong electric coupling. The equivalent circuit of a DTL cell is shown in Fig. 11. The coupling factor $k$ is in this case the ratio between the tube-to-wall and tube-to-tube capacitances $C/C_0$. A detailed analysis of the DTL equivalent circuit can be found in Ref. [2]. A DTL cavity is usually made of a large number of cells (up to more than 50 in a single tank), but because of the large coupling factor only the lowest modes can be observed, the others being hidden among the many different modes appearing at high frequencies.

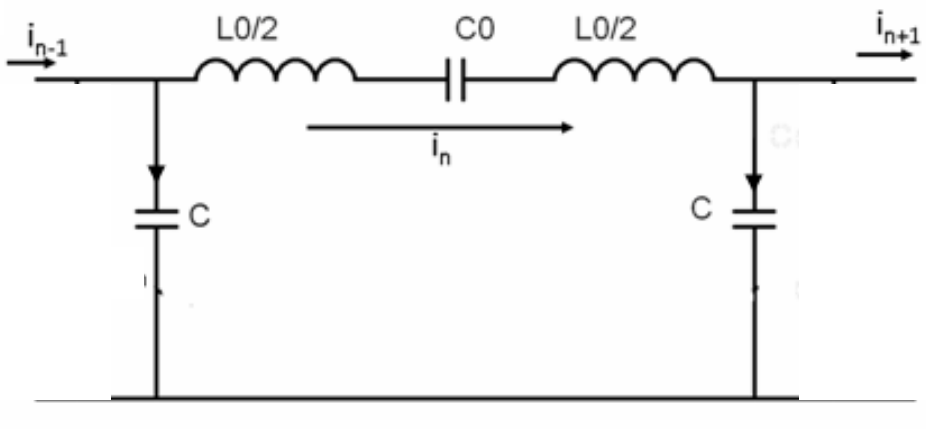

**Fig. 11:** Equivalent circuit of a DTL cell

The DTL is particularly suited to be used at low energies because the length of the cells (and of the drift tubes) can be easily adapted to the increasing velocity of the particle beam. We should observe that in the theoretical approach developed in Section 2, all relations depend only on the frequency and not on the inductance or capacitance of the single cells. If the capacitance and inductance from one cell to the other is changed keeping constant their product and therefore the cell frequency, all of the relations developed in Section 2 remain valid; the mode frequencies and the relative amplitudes in the cells will not change. The consequence is that the distance between gaps in a DTL can be easily adapted to the increasing beam velocity: if the length of the cells is progressively increased, keeping constant their frequency, the system will still behave in the usual way and the operating 0-mode will keep all of its properties. Tuning at the same frequency cells of different length is almost straightforward because increasing by the same proportion the cell and the drift tube lengths, the inductance will increase (longer cells) and the capacitance will decrease (larger gaps) by the same amount, and in a first approximation the variations will compensate keeping the frequency constant. Only minor adjustments to the gap lengths are required to compensate for second-order effects.

The possibility to adjust each individual cell length to the particle $\beta$ together with the option of easily inserting focusing quadrupoles in the structure makes the DTL an ideal structure for the initial acceleration in a proton linac, from energies of a few megaelectronvolts to some 50 MeV to 100 MeV. As an example, Fig. 12 shows a three-dimensional open view of the CERN Linac4 DTL, which will accelerate $H^-$ particles from 3 MeV to 50 MeV. The structure is divided into three individual 352.2 MHz resonators, for a total of 120 cells in a length of 19 m. The relativistic velocity increases from $\beta = 0.08$ to $\beta = 0.31$, and correspondingly the cell length $\beta\lambda$ increases by a factor of 3.9.

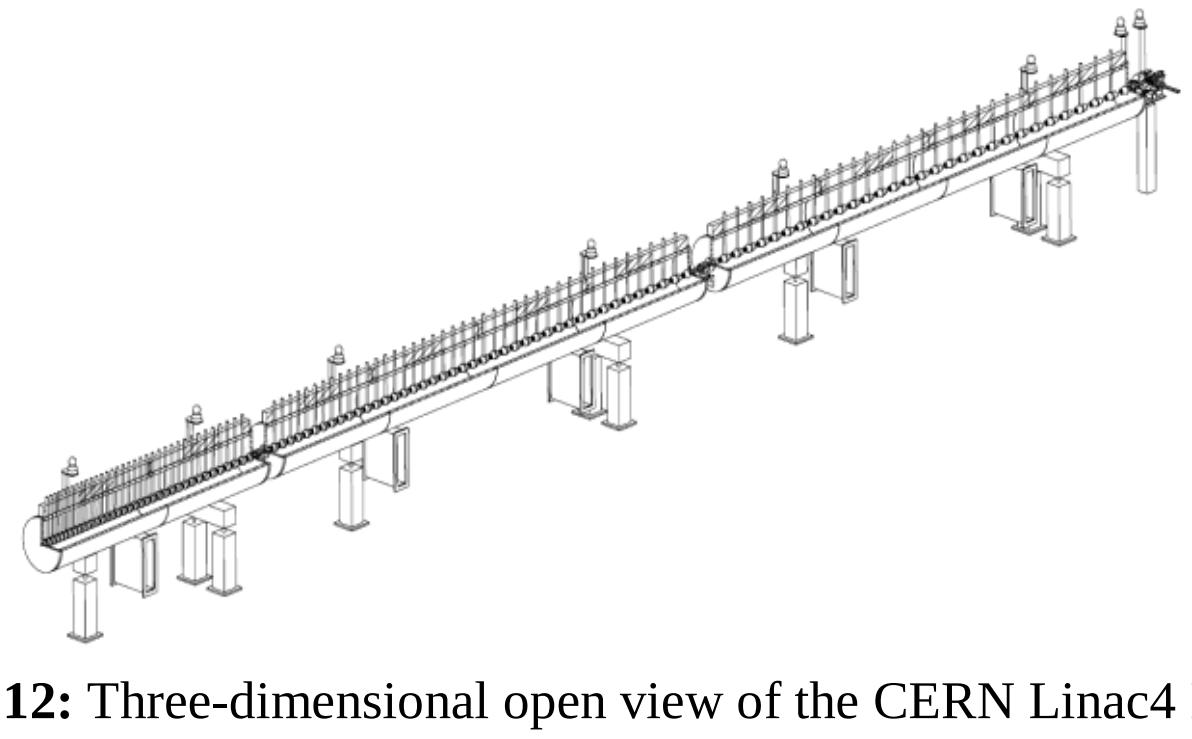

**Fig. 12:** Three-dimensional open view of the CERN Linac4 DTL

## 4 π-mode structures: PI-Mode Structure and elliptical cavities

Structures operating in the π mode are widely used, in particular in the magnetic coupled normal-conducting version and in the electric-coupled superconducting version. The coupling is provided by a slot on the external wall in the first case or by a sufficiently large opening on the axis for the latter. In both cases the cell length is kept constant inside short cavities made of a few (4 to 10, depending on the specific application) identical cells. Varying the cell length inside the cavities would complicate the design, because for π-mode structures not only the frequency but also the coupling factor depends on the cell length, and would considerably increase the construction cost. Therefore, π-mode structures are commonly used in the high-energy range of a linear accelerator for proton energies above 100 MeV, where the beam phase slippage is small.

As an example of normal-conducting π-mode structure, Fig. 13 shows the PI-Mode Structure (PIMS) that is being built at CERN for Linac4. Resonating at 352.2 MHz, it will cover the energy range between 100 MeV and 160 MeV. The PIMS cavities are made of seven cells, coupled via two slots in the connecting wall (visible on the left of Fig. 13); the pairs of slots on the two sides of a cell are rotated by 90° to minimize second-neighbour couplings that could perturb the dispersion curve. The complete PIMS section is made of 12 7-cell cavities. While the cell length inside each cavity is constant, it increases from cavity to cavity, matching the increase in $\beta$.

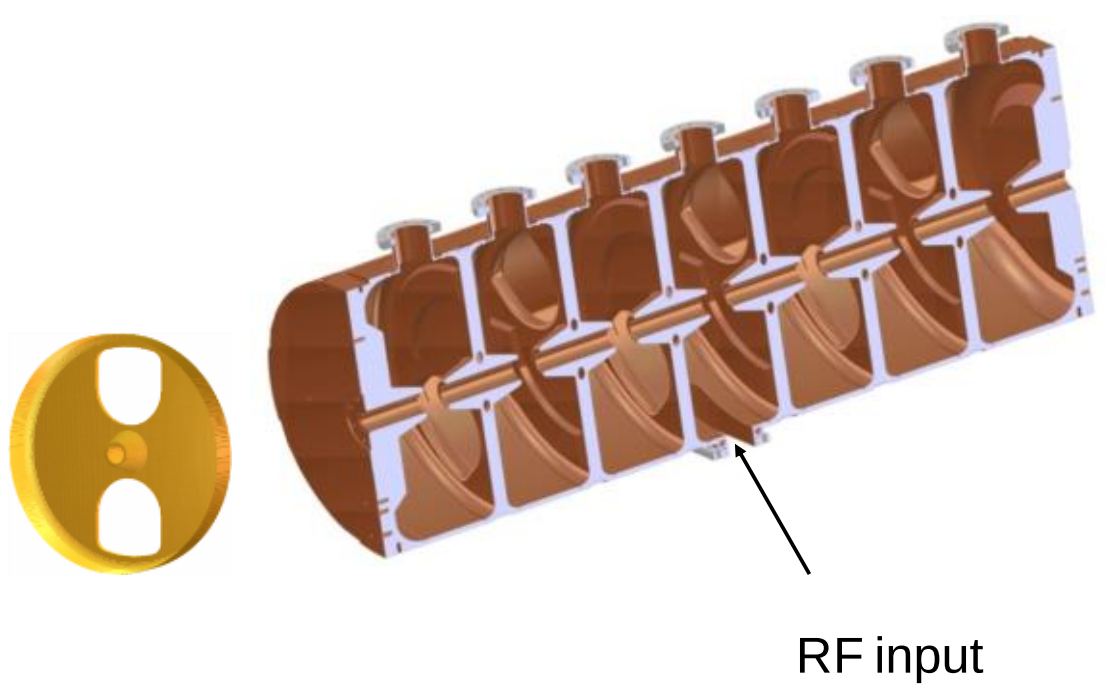

**Fig. 13:** The PIMS seven-cell cavity

Figure 14 shows a typical superconducting low-β cavity operating in the π mode. This particular cavity is made of five cells of identical length.

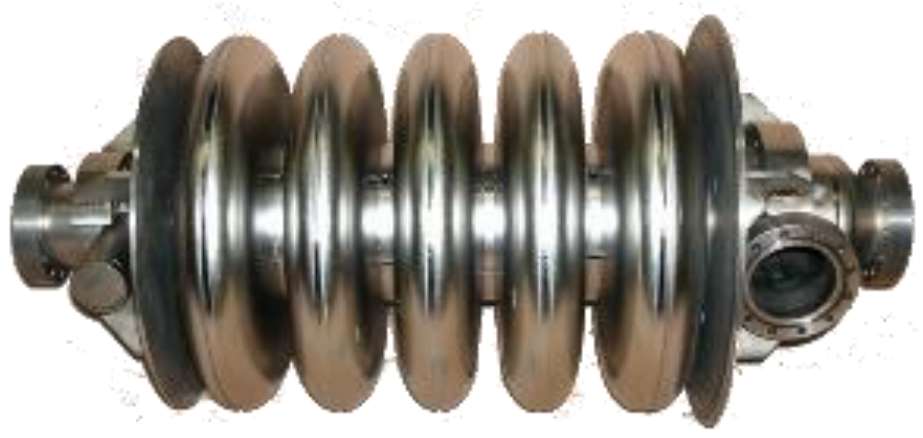

**Fig. 14:** A five-cell elliptical superconducting cavity

## 5 π/2-mode structures

To be able to operate very long chains of cells there is a particular interest in π/2-mode operation, although this mode has lower acceleration efficiency than 0 or π modes. This mode allows stable operation of long chains of coupled cells which can then be fed by single high-power RF sources, less expensive than many smaller power units.

Looking at Fig. 5, we see that the bandwidth of a coupled system is only proportional to the coupling factor and independent of the number of cells. Therefore, if we have a large number of cells and a large number of modes on the dispersion curve the modes will be very close in frequency to each other; this is particularly true for the 0 and $\pi$ modes that lie in a region of the curve where the derivative is small. The modes will remain separated because their $Q$ value is usually sufficiently high; however, an important consequence of having several other modes close in frequency to the operating mode is that the system becomes extremely sensitive to mechanical errors. Small deviations from the design frequency in some cells of the chain (coming, for example, from usual machining errors) would change the boundary conditions of the operating mode, forcing the system to introduce components from the adjacent modes to respect the new perturbed boundary conditions. These components are inversely proportional to the difference in frequency between operating and perturbing modes, making long structures more sensitive to errors than shorter ones.

In the $\pi/2$ mode instead, not only is the distance in frequency between the operating mode and the perturbing modes the largest, but their effect is compensated, making the chain of resonators virtually insensitive to the mechanical errors. The reason for this compensation is that the components from perturbing modes add up to the field distribution of the operating mode with a sign, positive or negative depending whether they are higher or lower in frequency than the operating mode. Observing that the modes on the two sides of the operating $\pi/2$ mode have the same field distributions in the cells that are excited (but different ones in the cells that are empty), an error in the chain of resonators will be compensated for by symmetric components of the modes higher or lower in frequency; these "perturbed" components will come with opposite sign and will cancel each other. In principle, a $\pi/2$-mode structure can be totally insensitive to errors; in practice, this requires a perfect symmetry of the perturbing modes around the operating one that is usually difficult to achieve. Reduction of the error sensitivity between a factor of 10 and a factor of 100 when going from a 0 or $\pi$ mode to a $\pi/2$ mode is usually considered as satisfactory.

The best known $\pi/2$ structure is the Side-Coupled Linac (SCL) structure (Fig. 15) developed at the Los Alamos National Laboratories in the 1960s. Here the coupling is magnetic, through slots on the cell walls, and the coupling cells are moved away from the beam axis and placed symmetrically on both sides of the chain of accelerating cells. The result is that from the electromagnetic point of view, the structure operates in the $\pi/2$ mode providing stabilization of the field, whereas the beam travelling on the axis sees the typical field distribution of a $\pi$ mode with maximum acceleration. Side-coupled structures are used at high energy and high RF frequency (from about 700 MHz), where high-power klystrons provide an economical way to feed a large number of cells, for which operation in 0 or $\pi$ mode would be impossible because of the strong sensitivity to mechanical errors.

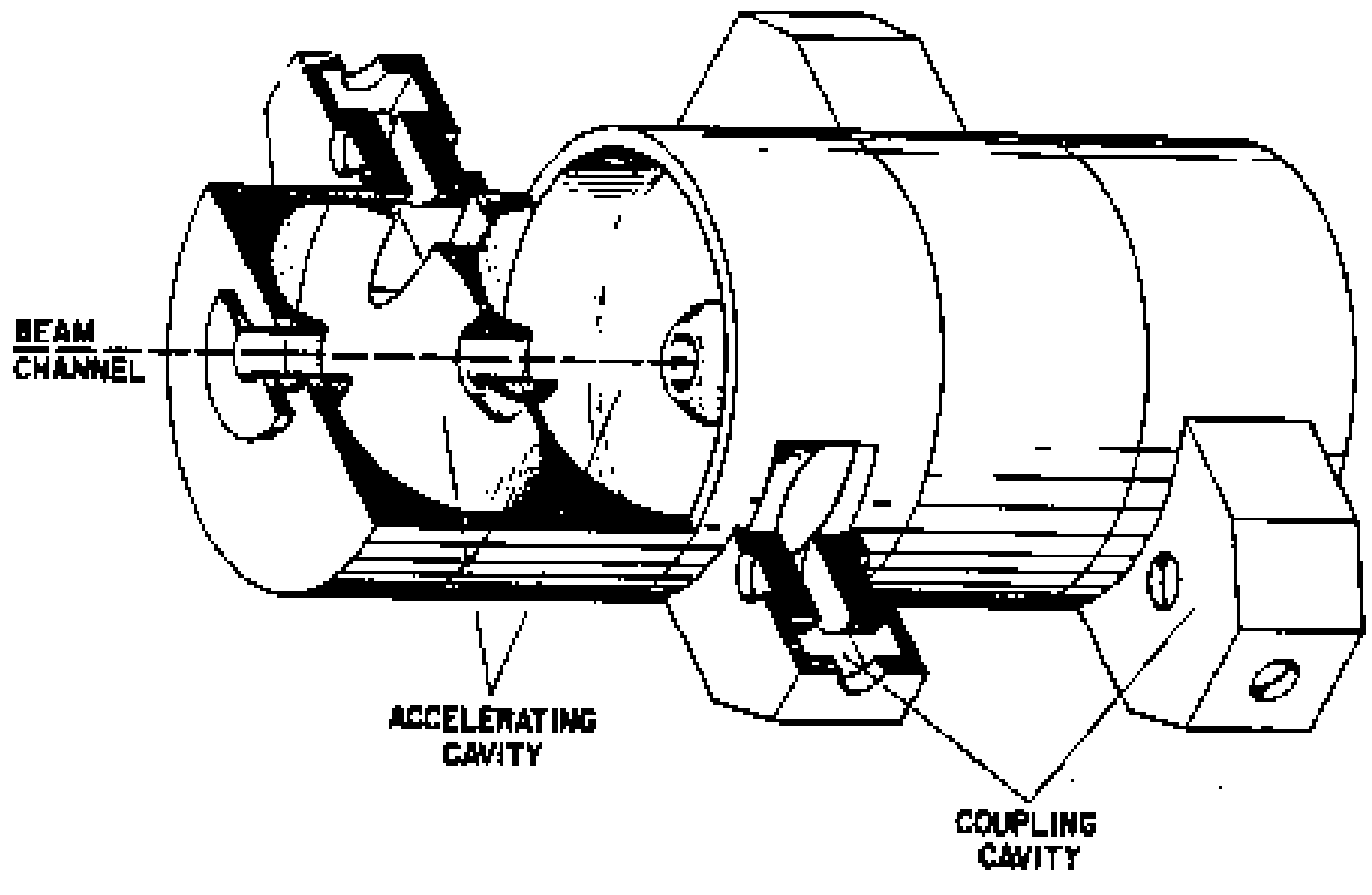

**Fig. 15:** The SCL structure [3]

## 6 Other accelerating structures: comparison of shunt impedances

We have so far considered the three main groups of coupled-cell accelerating structures, 0 mode, π mode and π/2 mode, and described the main representative for each category, the DTL, the PIMS and the SCL, respectively. Many more structures are used in linacs, however, which van be variants of these main families, can mix properties of different families or can even be based on different operating modes and a different approach. Among the most known "alternative" structures are the Separated-DTL (SDTL), a variant of the DTL without quadrupoles in the drift tubes, the Annular Coupled Structure (ACS), a π/2-mode structure with a different coupling cell geometry from the SCL, the Cell-Coupled DTL (CCDTL), a structure mixing a 0-mode with a π/2-mode operation, etc.

A particular category of linac structures is based on operation on a TE mode instead of the usual TM mode. The TE modes are called H modes in the German literature and these structures are usually referred as "H-mode" structures. A TE mode in principle has electric field only in the transverse direction and therefore cannot be used for acceleration. If drift tubes are placed in the structure connected alternatively to two sides of the resonator, however, the electric field of the TE11 mode can be forced in the longitudinal direction between the drift tubes and thus be able to provide acceleration. These are the so-called "IH structures". In a similar way, if the supports of the drift tubes are placed alternatively on the two transverse axes of the accelerating structure the TE21 mode can be forced to have a longitudinal electric field between the drift tubes: this is the "CH structure". IH and CH are very compact in terms of transverse dimensions; this is an advantage for low frequencies, but makes their construction difficult if high frequencies are required.

The choice of the most appropriate accelerating structure for a given project is very complex, being based on the comparison of many parameters. One of the most important figures of merit used for the selection of the accelerating structure is the shunt impedance, which represent the efficiency of a RF cavity in converting RF power into voltage across a gap. This is defined as

$$Z = \frac{V_0^2}{P}$$

with $V_0$ the peak RF voltage in a gap and $P$ the RF power dissipated on the cavity walls to establish the voltage $V_0$. When the reference is to the effective voltage seen by a particle crossing the gap at velocity $\beta$c, we define the effective shunt impedance as

$$ZT^2 = \frac{(V_0 T)^2}{P}$$

with $T$ the transit time factor of the particle crossing the gap (ratio of voltage seen by the particle during the crossing over maximum voltage available). If the structure has many gaps, we can refer to the shunt impedance per unit length, usually expressed in megaohms per metre. It must be noted that here we use the "linac" definition, considering the shunt impedance as a sort of efficiency, i.e. a ratio between useful work (the voltage available to the beam, which is proportional to the energy gained by a particle, squared for dimensional reasons) and the energy (power in this case) required to obtain it. If instead we start from the consideration that the shunt impedance is the equivalent resistance in the parallel equivalent circuit of a cavity resonator, we need to add a factor of two at the denominator of the previous relations. This is the "circuit" or "RF" definition of shunt impedance.

RF power is expensive, and the goal of every designer of normal-conducting accelerating structures is to maximize the shunt impedance, which depends on the mode used for acceleration, on the frequency and on the geometry of the structure. Other considerations come of course into play in the overall optimization; however, the shunt impedance remains one of the essential references for the structure designer.

Comparing structures in terms of shunt impedance is not easy, because the shunt impedance depends on the chosen frequency as well as on several design parameters related to the different projects. To make the comparison as objective as possible for several types of structures, a study made in 2008 by the EU-funded Joint Research Activity HIPPI (High-Intensity Pulsed Power Injectors) derived the shunt impedance curves presented in Fig. 16, comparing eight different designs being studied in three different European Laboratories [4].

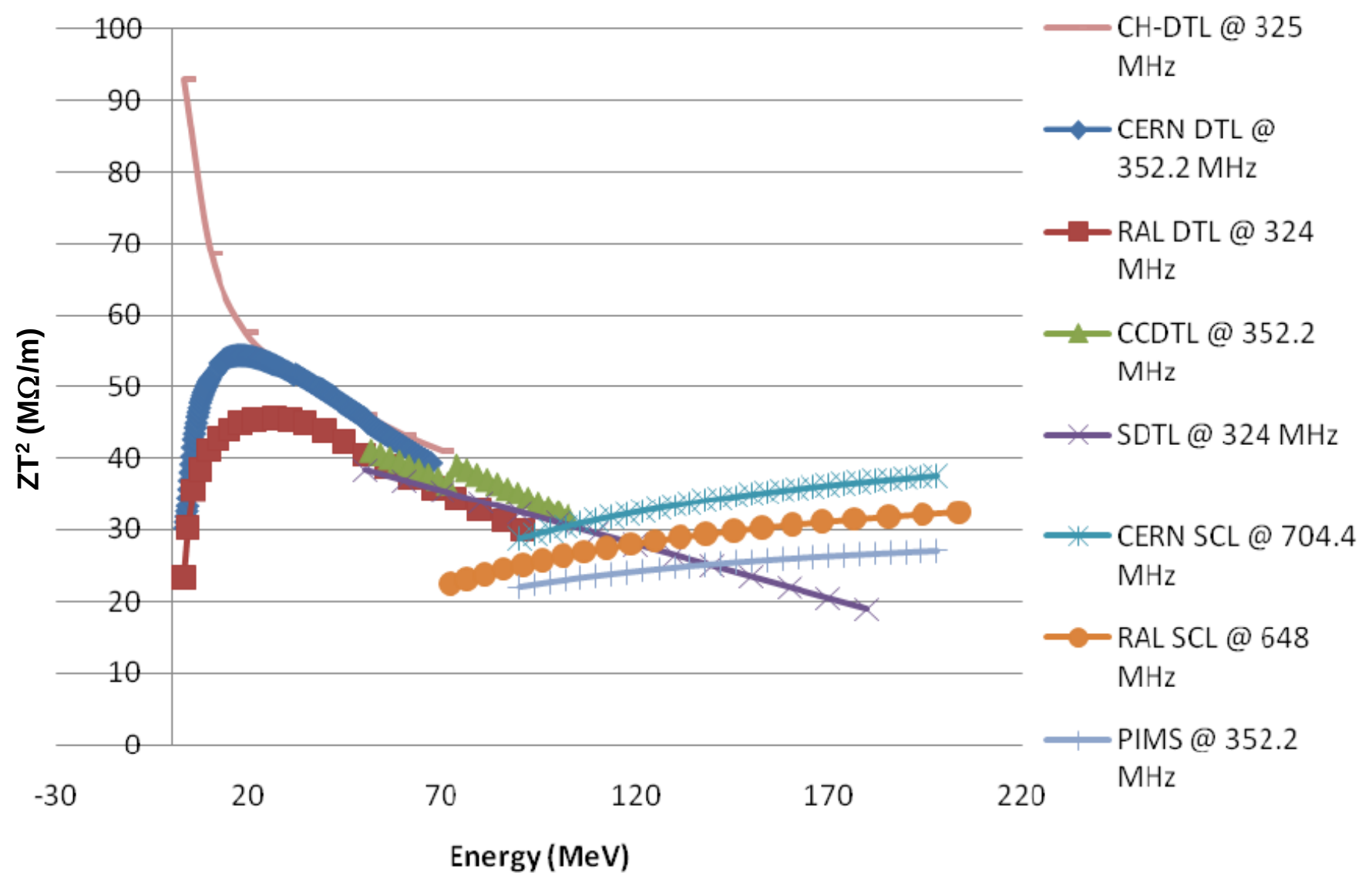

**Fig. 16:** Shunt impedance curves for different low-β structures

The values presented here correspond to simulations of effective shunt impedance per unit length corrected for the additional losses expected in the real case, for designs already optimized. The structures taken into consideration belong to two frequency ranges: the 324 MHz to 352 MHz range and its double harmonic, 648 MHz to 704 MHz. Higher operating frequencies have inherently higher shunt impedance; however, beam dynamics requirements in the first low-energy stages impose starting the acceleration at frequencies below about 400 MHz.

From the curves it appears that for all structures the shunt impedance has a more or less pronounced dependence on beam energy, due to the different distribution of RF currents and losses in cells of different length. Whereas 0-mode structures (DTL, but also the CCDTL in this context) have a maximum shunt impedance around 20 MeV to 30 MeV and then show a rapid decrease with energy, π-mode structures have a shunt impedance that is instead slightly increasing with energy, but starts from lower values than 0-mode structures. A natural transition point between these two types of structures would be around 100 MeV. For π-mode structures, remaining at the basic RF frequency leads to about 25 % lower shunt impedance than doubling the frequency (comparing CERN SCL and PIMS curves). Different considerations apply to H-mode structures; the CH considered in this comparison has by far the highest shunt impedance below 20 MeV. Above, its behaviour is similar to that of TM 0-mode structures.

## 7 Low-β superconducting structures

For superconducting structures, shunt impedance and power dissipation are not a concern, and the much lower RF power required allows using simpler and relatively inexpensive amplifiers. A separated-cavity configuration such as that shown in the bottom of Fig. 3 is therefore preferred for most superconducting linac applications at low energy, up to some 100 MeV to 150 MeV, where more operational flexibility is required and where the short cavity lengths allow having more quadrupoles per unit length, as required by beam focusing at low energy. At higher energies, superconducting linacs use multicell $\pi$-mode cavities such as that presented in Fig. 12.

We must, however, observe that only few low-$\beta$ linacs use single-gap cavities; even for superconducting structures, economic reasons suggest adopting structures with generally two or in some cases three or four gaps. The most widespread resonator used in particular for very low-beta heavy ion applications is the quarter-wavelength resonator (QWR, Fig. 17), sometimes declined in the half-wave resonator (HWR), when it is important to avoid even small dipole field components on the axis.

A resonator that has been proposed recently for several proton beam applications requiring operation at a large duty cycle, where superconductivity is an advantage, is the “spoke”. In this cavity the electric field across the gaps is generated by a magnetic field turning around some supports, the spokes. Its main advantages are the compact dimensions and the relative insensitivity to mechanical vibrations. Similarly, for intense proton or deuteron beams is proposed a superconducting version of the CH resonator. Some examples of these structures are presented in Fig. 17.

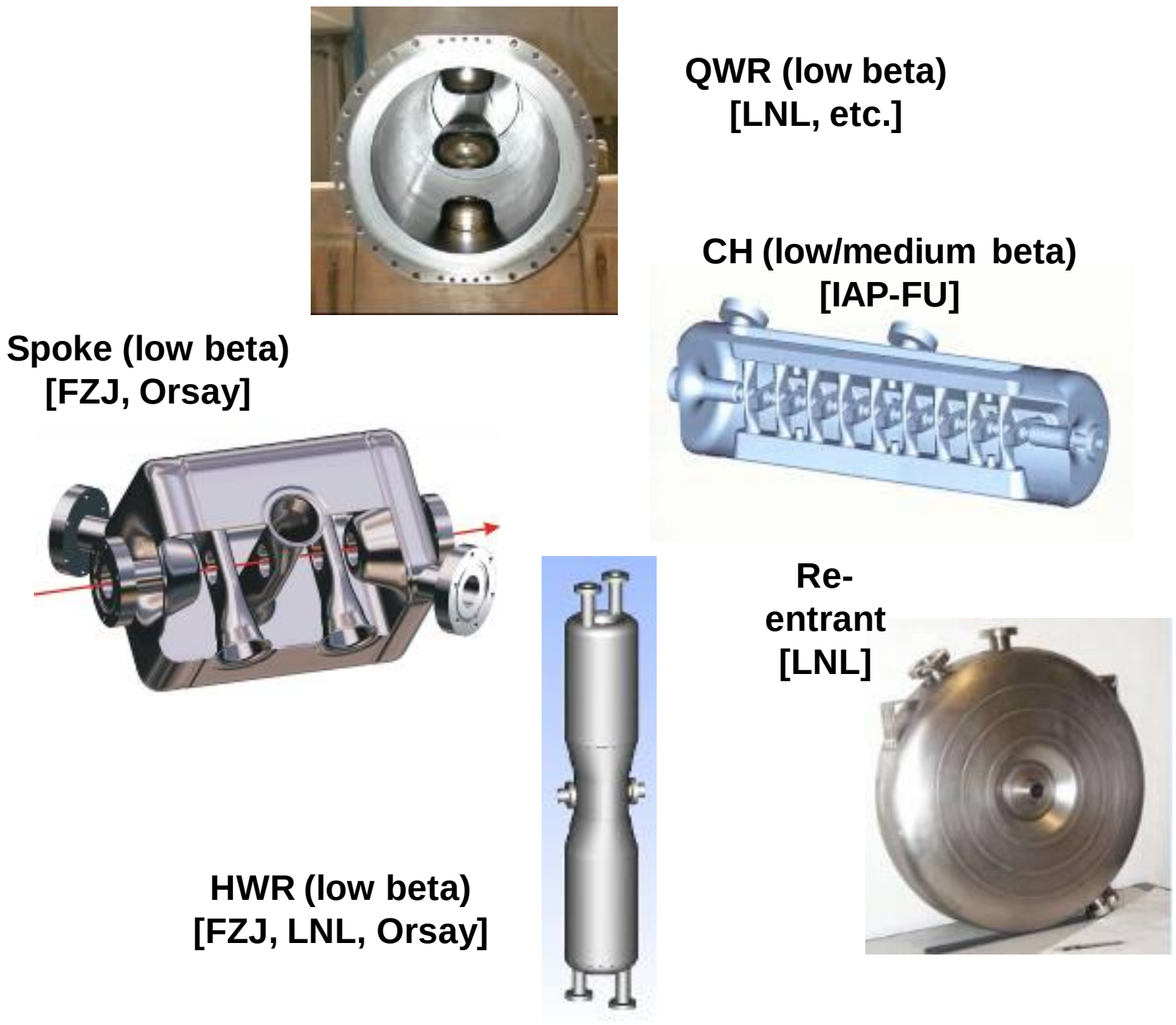

**Fig. 17:** Some examples of superconducting low-$\beta$ structures

# 8 Beam dynamics in linear accelerators

## 8.1 Longitudinal plane

We have seen that, to achieve the maximum acceleration, bunches of particles must be synchronous with the accelerating wave. This means that they have to be injected into the linac on a well-defined phase with respect to the accelerating sinusoidal field, and then they need to maintain this phase during the acceleration process. Linac beams are usually made of a large number of particles with a given spread in phase and in energy. If the injection phase corresponds to the crest of the wave ($\varphi = 0^{o}$ in the linac definition) for maximum acceleration, particles having slightly higher or lower phases will gain less energy. They will slowly lose synchronicity until they are lost.

In linacs, the same principle of phase stability holds as in synchrotrons: if the injected beam is not centred on the crest of the wave but around a slightly lower phase, a "synchronous phase" $\varphi_s$ whose typical values are between –20° and –30°, particles that are not on the central phase will oscillate around the synchronous phase during the acceleration process. The resulting longitudinal motion is confined, and the oscillation is represented by an elliptical motion of each particle in the longitudinal phase plane, i.e. the plane ($\Delta\varphi$, $\Delta W$) of phase and energy difference with respect to the synchronous particle. The relation between the synchronous phase in an accelerating sinusoidal field and the longitudinal phase plane is presented in Fig. 18.

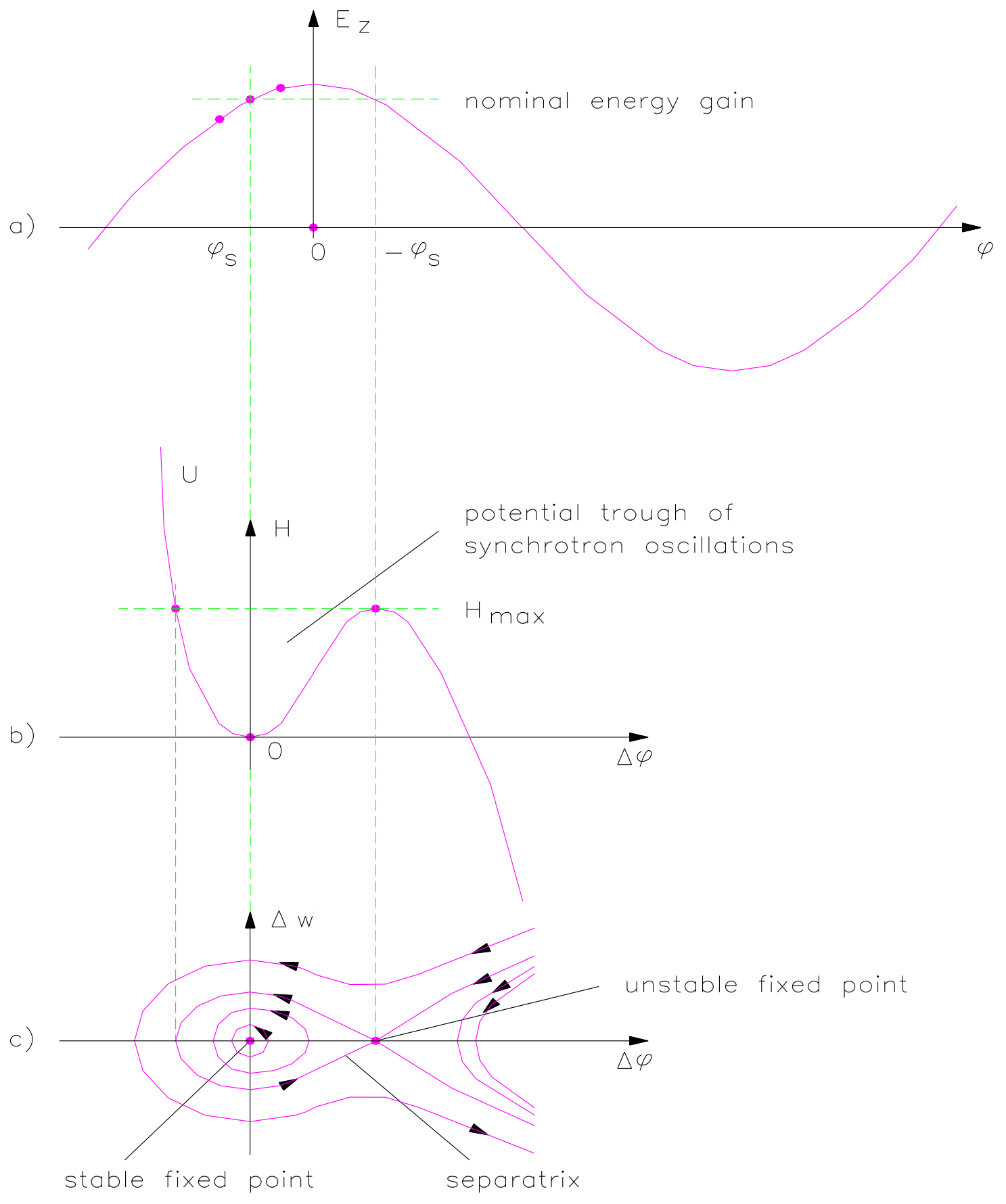

**Fig. 18:** Longitudinal motion of an ion beam

It is interesting to observe that the frequency of longitudinal oscillations, i.e. the number of oscillations in the longitudinal phase plane per unit time depends on the velocity of the beam. A simple approximate formula for the frequency of small oscillations $\omega_l$ can be found for example in Ref. [5]:

$$\omega_l^{\,2} = \omega_0^{\,2}\,\frac{qE_0T\sin(-\varphi)\lambda}{2\pi\, mc^2\beta\gamma^3}$$

Here $\omega_0$ and $\lambda$ are the RF frequency and wavelength, $E_0T$ is the effective accelerating gradient and $\varphi$ is the synchronous phase. The oscillation frequency is proportional to $1/\beta\gamma^3$: when the beam becomes relativistic, the oscillation frequency decreases rapidly. At the limit of $\beta\gamma^3 >> 1$, the oscillations will stop and the beam is practically "frozen" in phase and in energy with respect to the synchronous particle. For example, in a proton linac $1/\beta\gamma^3$ and correspondingly $\omega_l$ can decrease by two or three orders of magnitude from the beginning of the acceleration to the high-energy section.

Another important relativistic effect for ion beams is the "phase damping", the shortening of bunch length in the longitudinal plane. This can be understood considering that, as the beam becomes more relativistic, its length in *z* seen by an external observer will contract due to relativity. A precise relativistic calculation shows that the phase damping is proportional to $1/(\beta\gamma)^{3/4}$:

$$\Delta\varphi = \frac{const}{(\beta\gamma)^{3/4}}$$

When a beam becomes relativistic, not only do its longitudinal oscillations slow down, but the bunch will also compact around the centre particle.

## 8.2 Transverse plane

Transversally in a linac, the beam will be subject to an external focusing force, provided by an array of quadrupoles or solenoids. This force has to counteract the defocusing forces that either develop inside the particle beam or come from the interaction with the accelerating field. The main defocusing contributions come from space charge forces and from RF defocusing.

### *8.2.1 Space charge forces*

They represent the Coulomb repulsion inside the bunch between particles of the same sign. In the case of high-intensity linacs at low energy, space charge forces are one of the main design concerns. At relativistic velocity, however, the space charge repulsion starts to be compensated by the attraction due to the magnetic field generated by the beam, and finally disappears at the limit $v = c$. Space charge forces can be calculated only for very simple cases, such as that (Fig. 19) of an infinitely long cylindrical bunch with density $n(r)$ travelling at velocity $v$ .In this case, the electric and magnetic fields active on a particle at distance $r$ from the axis can be written as

$$E_r = \frac{e}{2\pi\varepsilon\, r}\int_0^r n(r)\, r\, dr \qquad\qquad E_r = \frac{e}{2\pi\varepsilon\, r}\int_0^r n(r)\, r\, dr$$

The resulting overall force acting on a particle in the bunch is orientated in the radial direction, and has intensity

$$F = e(E_r - vB_\varphi) = eE_r(1-\frac{v^2}{c^2}) = eE_r(1-\beta^2) = \frac{eE_r}{\gamma^2}$$

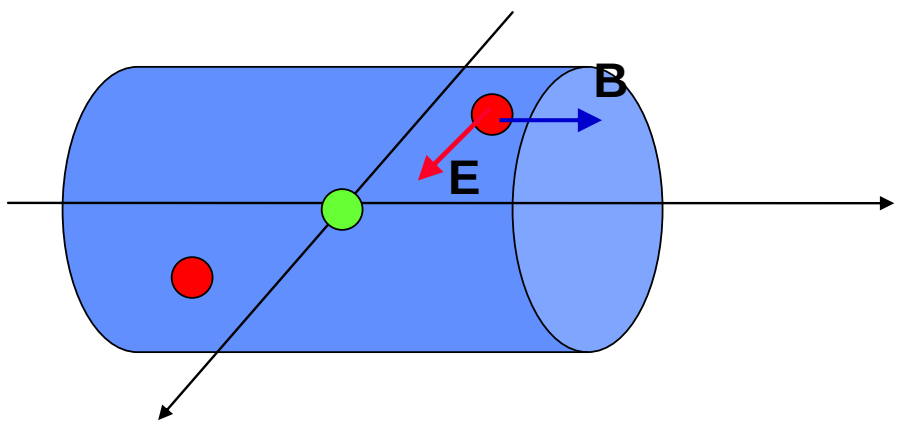

**Fig. 19:** Forces acting on a particle inside an infinitely long bunch

The overall space charge force is then proportional to $1/\gamma^2$ and will disappear for $\gamma \to \infty$.

### *8.2.2 RF defocusing forces*

The RF defocusing is the transverse defocusing experienced by a particle that crosses an accelerating gap on a longitudinally focusing RF phase. We have seen in the previous section that for longitudinal stability the beam will cross the gap when the field is increasing ($\varphi_s < 0$). Figure 20 shows a schematic configuration of the electric field in an accelerating gap. In correspondence to the entry and exit openings for the beam, the electric field has a transverse component, focusing at the entrance to the gap and defocusing at the exit, proportional to the distance from the axis. Because the field is increasing when the beam crosses the gap, the defocusing effect will be stronger than the focusing effect, and the net result will be a defocusing force proportional to the time required by the beam to cross the gap. A Lorentz transformation from the laboratory frame to the frame of the particles of the electric and magnetic field forces acting on a particle allows calculating the radial momentum impulse per period. Carrying out this calculation, one can find:

$$\Delta p_r = -\frac{\pi e E_0 T L r \sin\varphi}{c \beta^2 \gamma^2 \lambda}$$

Again, this effect is proportional to $1/\gamma^2$, and will disappear at high beam velocity.

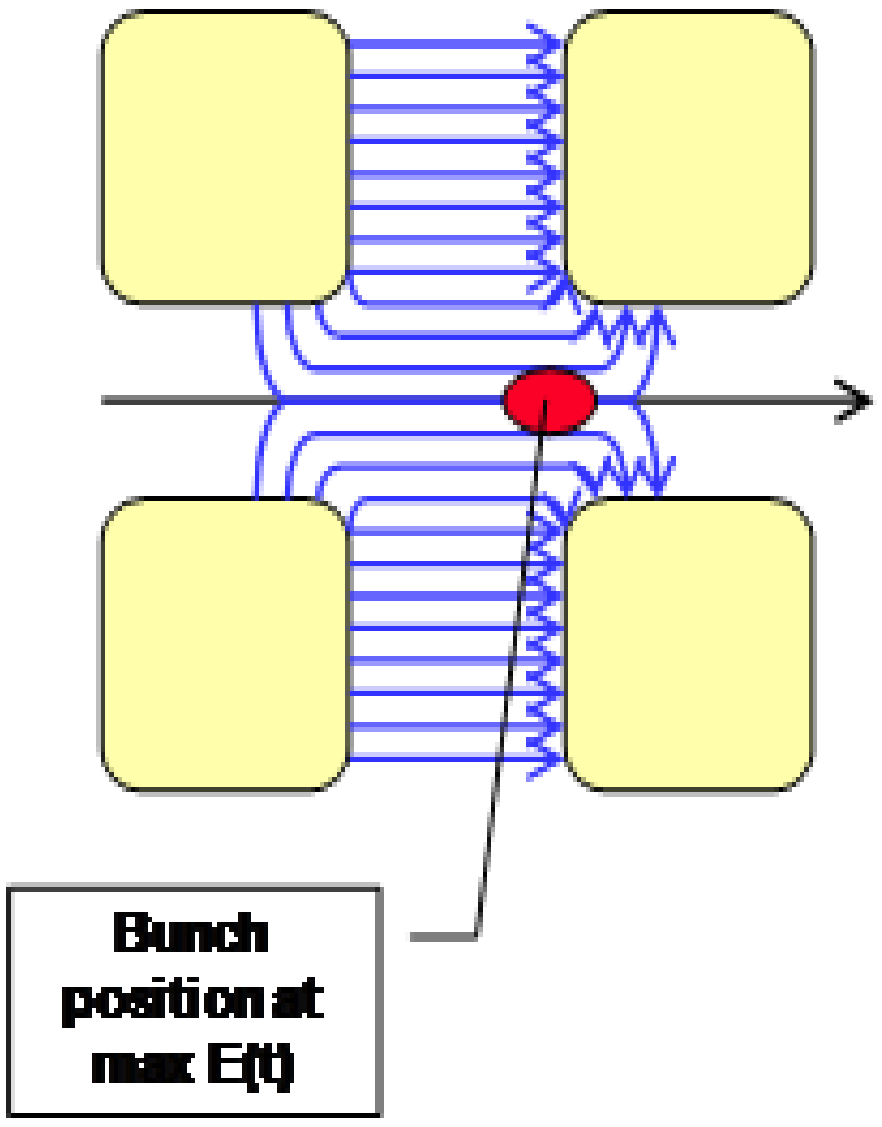

**Fig. 20:** Electric field line configuration around a gap and position of the bunch at maximum field

### 8.3 Transverse equilibrium

We have seen that in particular at low energies strong transverse defocusing forces act on the beam; to transport it with minimum particle loss we need to compensate for the defocusing forces with focusing forces, using standard alternating gradient focusing provided by quadrupoles along the beam line. As we have seen, a linac provides its acceleration with a series of accelerating structures; the standard focusing solution consists (Fig. 21) of alternating accelerating sections with focusing sections made of one quadrupole (singlet focusing), two quadrupoles (doublet focusing) or three quadrupoles (triplet focusing). In this way, we can immediately define the *focusing period* of the linac, corresponding to the length after which the structure repeats itself. It is important to observe that because the accelerating sections have to match the increasing beam velocity, the accelerating structures can have increasing lengths and therefore the basic focusing period does not necessarily have a constant length; however, in this case the travel time of the beam within a focusing period remains constant. The maximum length of the accelerating structure between the focusing elements depends on the beam energy, as we will see in the following; it goes from only one gap in the DTL to one or more structures containing many gaps at high energies.

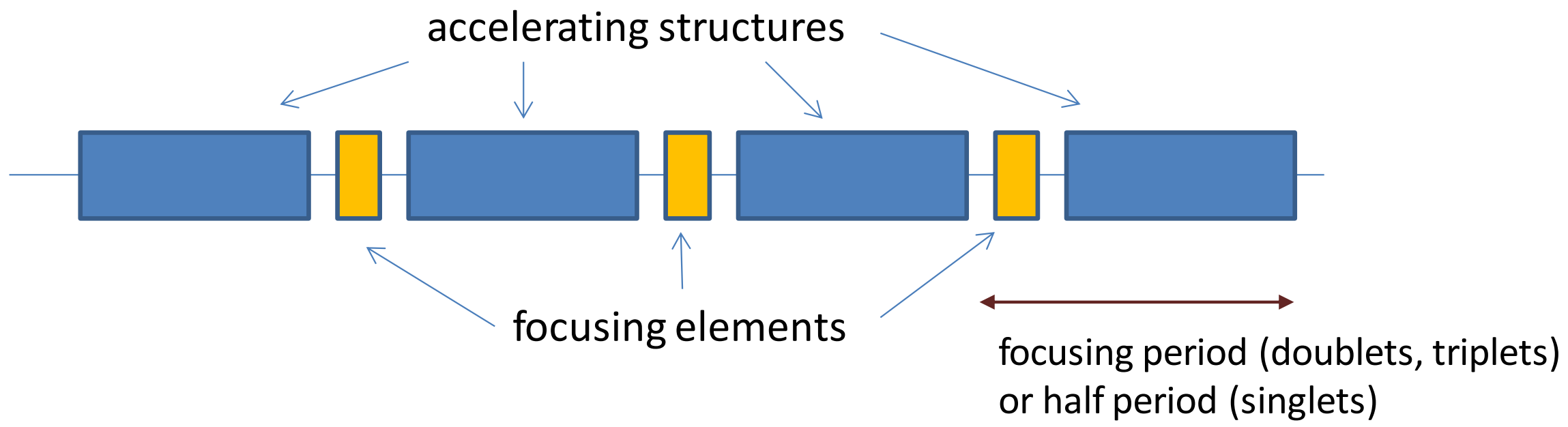

**Fig. 21:** Standard linac section

The beam must be transversally in equilibrium between the external focusing forces and the internal defocusing forces; the equilibrium will necessarily be dynamic, resulting in an oscillation in time and in space of the beam parameters with a frequency that will depend on the difference between focusing and defocusing forces. The oscillation can be as usual decomposed in two independent oscillations with the same frequency in the transverse planes $x$ and $y$, the reference oscillating parameter being as usual the maximum beam radius. The oscillation is characterized by its frequency; instead of defining it with respect to time it is convenient to define the oscillation frequency in terms of the phase advance $\sigma_t$, the increase in the phase of the oscillation over one focusing period of the structure, which is usually constant or changes only smoothly over a linac section. Alternatively, the phase advance per unit length $k_t$ can be used. If $L$ is the period length of the focusing structure, then $k_t = \sigma_t/L$. Modern beam dynamics simulation codes allow us to easily calculate the phase advance for a given focusing period, accelerating structure design and quadrupole gradient; it is important to select accurately this parameter to minimize beam loss and to avoid excessive transverse emittance growth. The first basic rule is that $\sigma_t$ should always be <90º, to avoid resonances that could lead to emittance growth and to beam loss; it should also be higher than some 20º, to avoid that the amplitude of the oscillations becomes too high and the beam size becomes too large.

We can find an approximate relation for the phase advance as a function of focusing and defocusing forces referred to a simple theoretical case. First of all, one has to limit the analysis to beam oscillations in a simple F0D0 quadrupole lattice (focusing–drift–defocusing-drift, corresponding to the "singlet" focusing) under smooth focusing approximation, i.e. averaging the localized effect of the focusing elements. Then, adding together the focusing and RF defocusing contributions to phase advance as derived for example in Ref. [5, Eq. (7.103)] and subtracting the space charge term as approximately calculated in the case of a uniform three-dimensional ellipsoidal bunch [5, Eq. (9.51)] we obtain for the phase advance per unit length:

$$k_t^2 = \left(\frac{\sigma_t}{N\beta\lambda}\right)^2 = \left(\frac{qGl}{2mc\beta\gamma}\right)^2 - \frac{\pi q E_0 T \sin(-\varphi)}{mc^2\lambda\beta^3\gamma^3} - \frac{3qI\lambda(1-f)}{8\pi\varepsilon_0 r_0^3 mc^3\beta^2\gamma^3}$$

Here $N\beta\lambda$ is the length of the focusing period in units of $\beta\lambda$. The first term on the right-hand side of the equation is the focusing component: $Gl$ is the quadrupole integrated gradient, expressed as product of gradient $G$ and length $l$ of the quadrupole. The second term is the RF defocusing: $E_0T$ is the effective accelerating gradient, $\lambda$ the RF wavelength and $\varphi$ the synchronous phase. For $\varphi < 0$, corresponding to longitudinal stability, $\sin(-\varphi)$ is positive and this term is negative, i.e. defocusing. The third term is the approximate space charge contribution: $I$ is the beam current, $f$ is an ellipsoid form factor ($0 < f < 1$) and $r_0$ is the average beam radius. The other parameters in the equation define the particle and medium properties (charge $q$, mass $m$, relativistic parameters $\beta$ and $\gamma$, free space permittivity $\varepsilon_0$).

This simple equation shows, although in an approximate simplified case, how the beam evolution in a linear accelerator depends on the delicate equilibrium between external focusing and internal defocusing forces. Real cases can be solved only numerically; however, the parametric dependence given by this equation remains valid, and allows us to determine how the beam dynamics will change with the particle $\beta$. At low velocities ($\beta << 1$, $\gamma \sim 1$) the defocusing terms are dominant. To keep the beam focused with a large enough phase advance per unit length one has to increase the integrated gradient $Gl$ and/or decrease the length of the focusing period $N\beta\lambda$, i.e. minimize the distance between focusing elements. This is for example the case of the radio-frequency quadrupole (RFQ), the structure of choice for low-energy ion beams (from $\beta \approx 0.01$ to $\beta \approx 0.1$). The RFQ provides a high focusing gradient by means of an electrostatic quadrupole field, with short cells at focusing period $\beta\lambda$. At higher energy, standard electromagnetic quadrupoles have a sufficiently high gradient and a structure alternating accelerating gaps and quadrupoles can be used. The standard structure used for protons above about 3 MeV energy is the DTL (Fig. 10), which presents a $2\beta\lambda$ focusing period when focusing and defocusing quadrupoles alternate inside the drift tubes (i.e. a F0D0 focusing). The maximum gradient achievable in the DTL quadrupoles together with the short focusing period allow keeping the phase advance in an acceptable range even for high current and high space charge beams; for example, the CERN Linac2 can accelerate a beam current up to 180 mA, the maximum achieved so far in a DTL. As an example of DTL beam dynamics design, Fig. 22 presents quadrupole gradients and the corresponding phase advance for the CERN Linac4 DTL design [6]. The corresponding oscillations of the beam envelope (in this case is plotted the maximum beam radius in $x$) for the given input matching conditions are shown in Fig. 23. As the phase advance decreases, the period of the oscillations becomes longer.

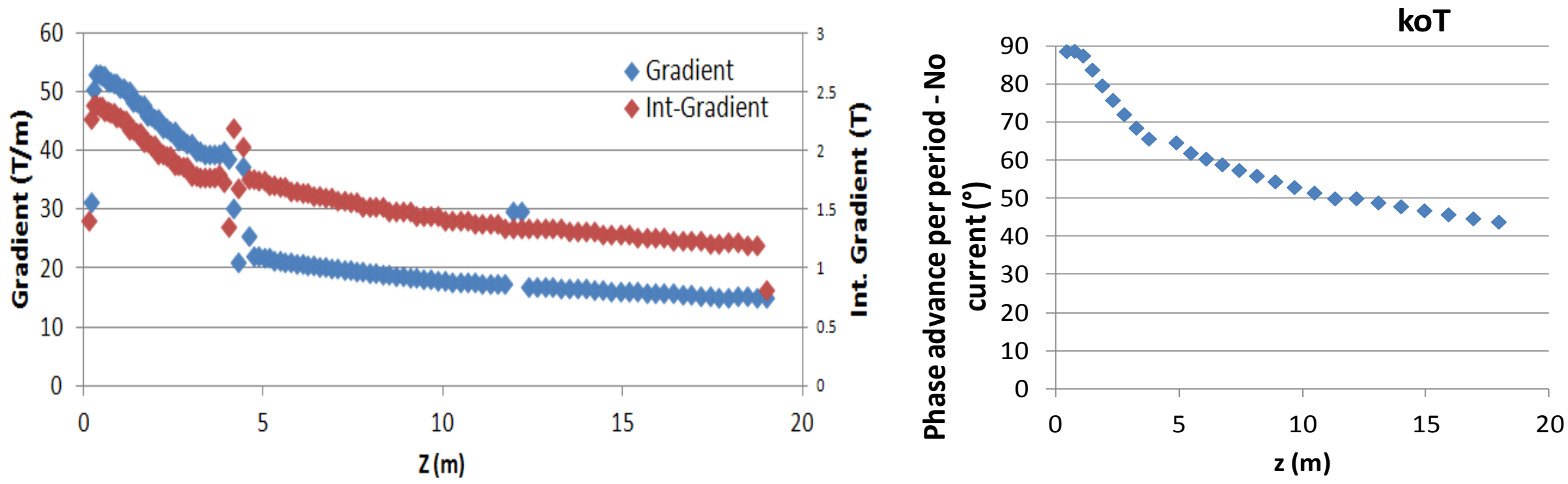

**Fig. 22:** Quadrupole gradients and corresponding phase advance for the Linac4 DTL design

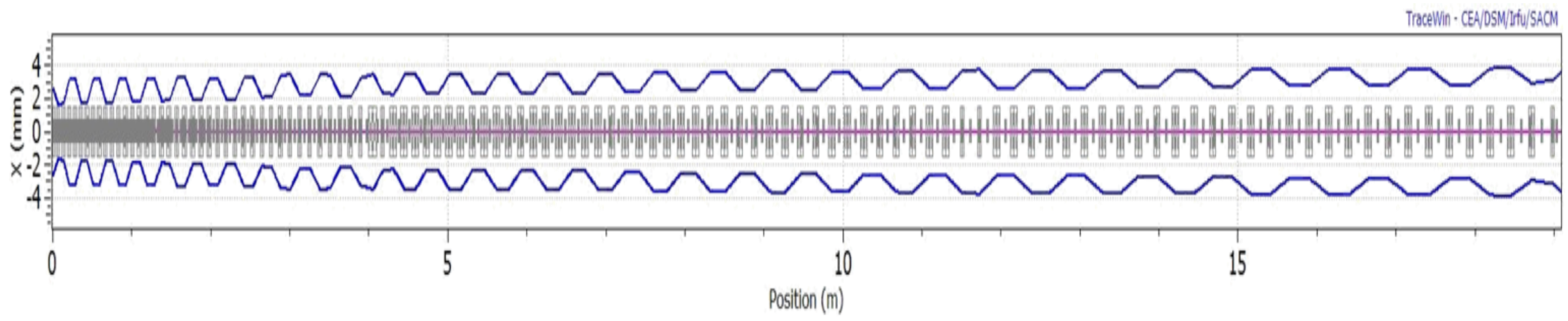

**Fig. 23:** Transverse root mean square beam envelope ($x$-plane) for the Linac4 DTL design

Going further up in energy, the defocusing terms (proportional to $1/\beta^3\gamma^3$ and $1/\beta^2\gamma^3$, respectively) decrease much faster than the focusing term (proportional to $1/\beta\gamma$). The focusing period can be increased, reducing the number of quadrupoles and simplifying the construction of the linac. Starting from energies between 50 MeV and 100 MeV modern proton linacs usually adopt multicell structures operating in $\pi$ mode spaced by focusing quadrupoles; these can be normal conducting as in the structure of Fig. 13 or superconducting as in that of Fig. 14. For example, in the CERN Linac4 design (90 keV to 160 MeV beam energy) the focusing period increases from $\beta\lambda$ in the RFQ to $15\beta\lambda$ in the last $\pi$-mode accelerating structure. The corresponding beam envelope is shown in Fig. 24. Heavy ions differ from protons for the fact that usually ion currents are small and the space charge term can be neglected; immediately after the RFQ, the focusing period can be some $5\beta\lambda$ to $10\beta\lambda$.

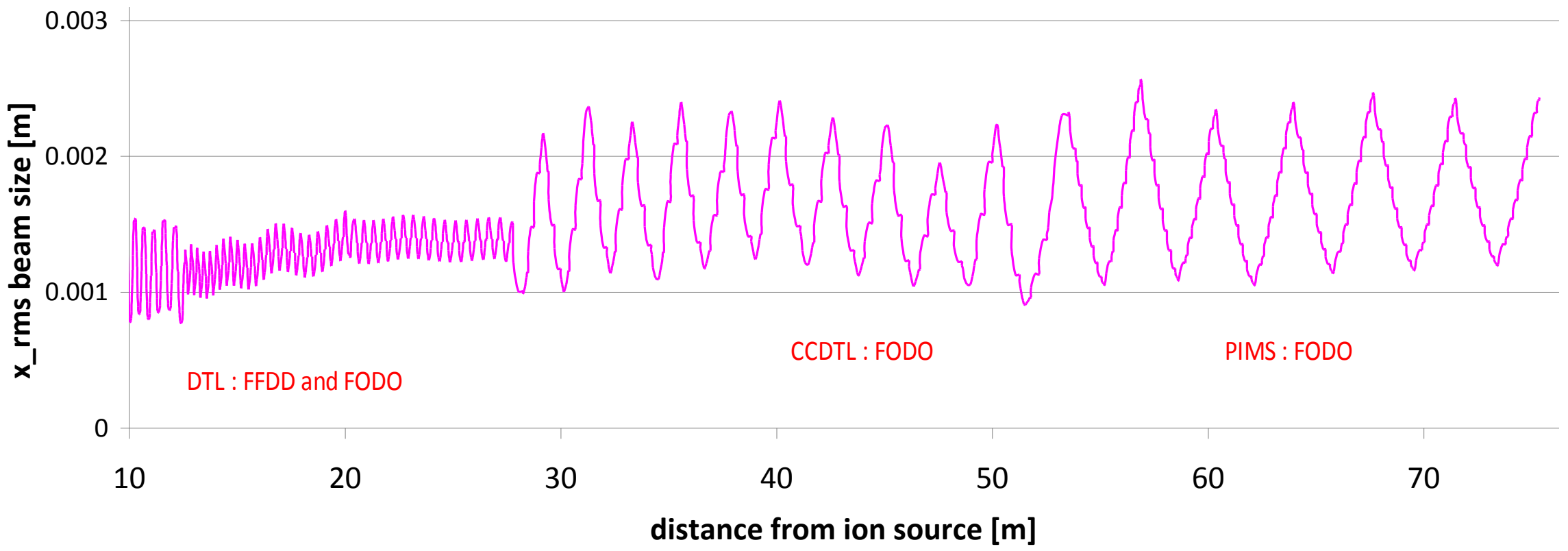

**Fig. 24:** Transverse root mean square beam envelope (*x*-plane) along the different Linac4 sections (3 MeV to 160 MeV)

## 9 Linac architecture

From the previous sections, we can derive two fundamental rules that define the basic architecture of any linear accelerator required to reach energies above a few megaelectronvolts:

(a) at low energy the linac cells have to be different in length to follow precisely the increase in beam velocity; at higher energy instead, the accelerating structures can be made out of sequences of identical cells, thus reducing the construction costs thanks to a higher standardization and to the use of longer vacuum structures;

(b) as the beam energy increases less focusing is required because of the reduction in space charge and RF defocusing; at higher energy the focusing length can increase, reducing the number and cost of the quadrupoles.

These two rules lead to the same consequence: to keep construction and operation costs low a linac must change the type of structure and focusing scheme with the increase of energy, going from expensive structures covering the low-energy range to more economical structures and focusing layouts for the high-energy range. A linac will be thus made of different sections covering different ranges of energy. After the ion source, the first accelerating structure will be a RFQ; this is the only linac structure that can provide the strong focusing forces needed to compensate for the space charge, which is very high at low energy and constitutes the main beam current limitation in linacs. The RFQ is an expensive structure which is not particularly efficient in using the RF power; for this reason, for an energy of between 2 MeV and 3 MeV it becomes convenient to go from the RFQ to another type of structure. Most linac projects use a DTL for the following section; the DTL integrates in a single RF cavity cells of increasing length matching precisely the increase in beam velocity and quadrupoles in every cell providing a strong and uniform focusing. As an alternative to the DTL, projects required to operate at very high duty cycle (close to or at CW) and/or to provide acceleration of different ion types can find an economic interest in using short superconducting structures, e.g. with two gaps at fixed distance. Both a normal-conducting DTL and a sequence of short superconducting cavities are quite expensive, and at energies of some tens of megaelectronvolts (usually between 50 MeV and 100 MeV) it is convenient to go to structures with many identical cells interleaved with quadrupoles; whereas at intermediate energies (50 MeV to 200 MeV) special structures are required, at higher energies long standardized modules of identical cells, often superconducting, can be used. Focusing is provided by quadrupoles placed between the modules.

As an example, Fig. 25 shows the layout of the CERN Linac4. This linac will cover the energy range between 45 keV (extraction from the ion source) and 160 MeV with four different types of accelerating structures, characterized by an increasing number of identical cells per accelerating structure and an increasing length of the focusing period.

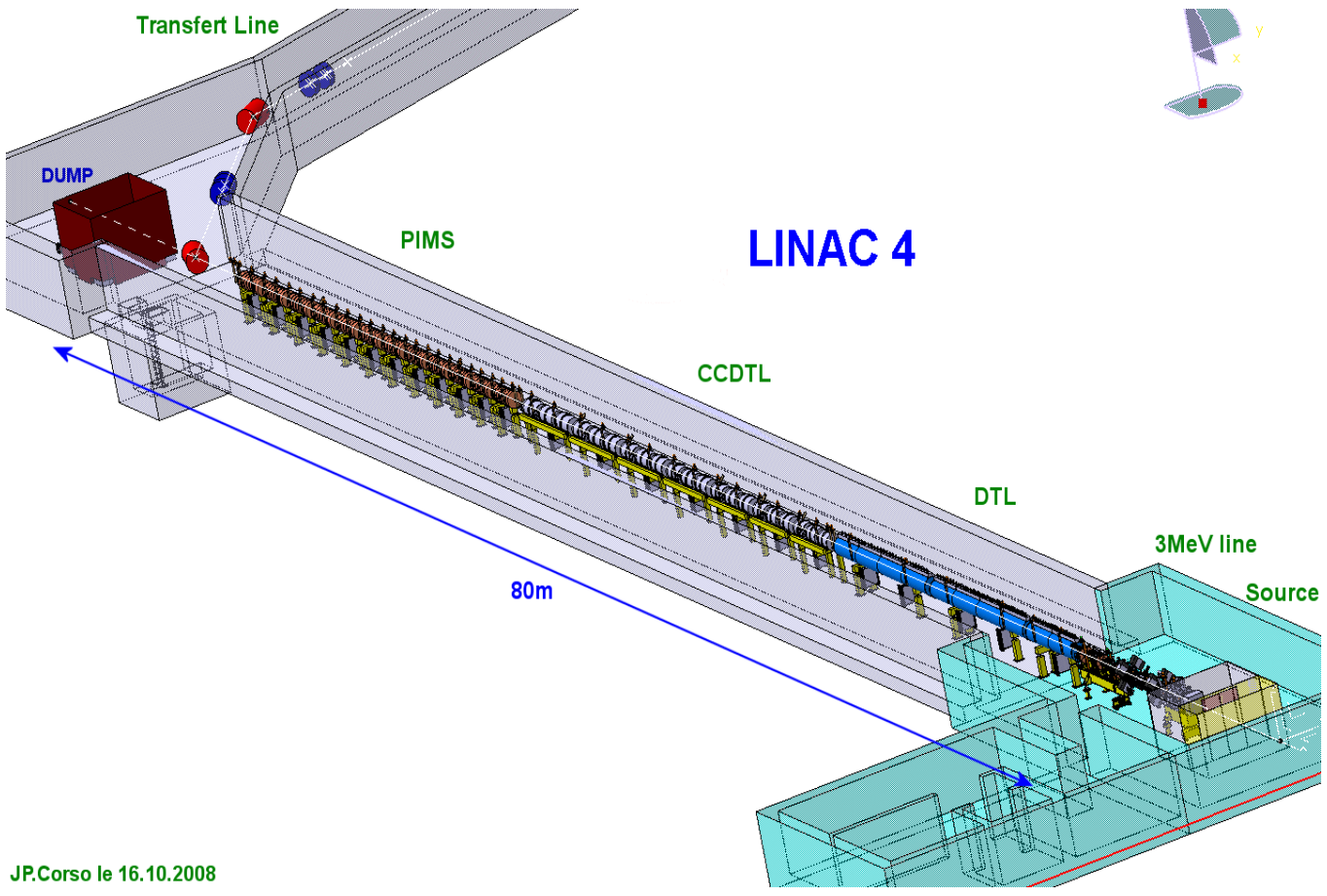

| Section | Output Energy [MeV] | Number of consecutive identical cells | Distance between quadrupoles |
|---|---|---|---|
| RFQ | 3 | 1 | - |
| DTL | 50 | 1 | βλ |
| CCDTL | 100 | 3 | 4 βλ |
| PIMS | 160 | 7 | 5 βλ |

**Fig. 25:** Layout of Linac4 under construction at CERN

### 9.1 The choice of the frequency

The choice of the appropriate frequency (or sequence of frequencies) for a linac has to take into account several factors coming from mechanical, RF and beam dynamics considerations related to the different scaling with frequency of the main linac parameters. First of all, accelerator dimensions and cell length are proportional to the wavelength λ: higher frequencies will result in smaller accelerating structures, requiring less copper or steel. Machining tolerances, however, scale as well as λ: smaller structures require tighter tolerances that are more expensive to achieve. From the RF point of view, higher frequencies are also preferable, because both the shunt impedance and the maximum surface electric field scale approximately as $\sqrt{f}$ .[4] In Section 8, however, we have seen that the RF defocusing scales with 1/λ, becoming excessively high at high frequencies; RF defocusing and cell length in the RFQ usually define the maximum frequency that can be used in the initial section of a linac. A summary of the dependence with the frequency for the different linac parameters is given in Table 1.

**Table 1:** Scaling with frequency of some basic linear accelerator parameters

RF defocusing $\sim f$
Cell length (~βλ) $\sim 1/f$
Peak electric field $\sim\sqrt{f}$
RF power efficiency (shunt impedance) $\sim\sqrt{f}$
Accelerator structure dimensions $\sim 1/f$
Machining tolerances $\sim 1/f$

---

[4] It should be noted that the scaling of the maximum field with the square root of the frequency is valid only approximately and for frequencies below about 10 GHz. The shunt impedance scales as the square root of the frequency in the case of a symmetric scaling of the cavity dimensions; if the beam aperture is kept constant, the dependence is smaller.

A first analysis of the frequency dependence indicates that high frequencies are economically convenient: the linac is smaller, makes use of less RF power and can reach a higher accelerating field. Limitations to the frequency come from the mechanical construction costs that depend critically on the required tolerances and on the RF defocusing in the RFQ; for these reasons, modern linac designs tend to start with a basic frequency in the RFQ and then to double it as soon as the cells become longer and the RF defocusing decreases. The availability and cost of the RF power sources is also an essential element in the choice of the frequency, and has to be considered carefully in particular when multiple frequencies are used.

All of these requirements result in some standardization in the frequencies commonly used in linacs: while in the past proton RFQs used to have frequencies around 200 MHz, nowadays frequencies in the range 325 MHz to 405 MHz are the usual standard. In the following sections, normal conducting or superconducting, a frequency jump by a factor of 2 is often applied, reaching the range 700 MHz to 800 MHz. A jump by a factor of 4 from 325 MHz allows reaching the International Linear Collider frequency of 1.3 GHz, thus making the use of protons of RF technology developed for electron linacs possible.

## 9.2 Superconductivity and the warm–cold transition

A constant in the architecture of modern high-energy linacs is the use of superconductivity at high energy. The advantages of superconducting accelerating structures are evident: a much smaller RF system needs to deliver only the power directly going to the beam, a large beam aperture allows for a lower beam loss (although the particles in the beam halo could be transported in the superconducting section and then lost in the following beam transport line) and finally the operating electricity costs are lower than for a normal-conducting linac, a feature particularly important given the present concerns for energy saving. The energy argument is particularly important for high duty cycle machines, where high power efficiency is an important requirement, and becomes proportionally less important for low duty machines where the beam power is a minor fraction of the power required by the machine.

In defining an optimum linac design, however, some peculiar characteristics of superconducting systems have to be considered. A superconducting system needs a large cryogenic installation requiring a significant amount of power; for a low duty linac this can be dominated by the static losses required for keeping the system at cryogenic temperature, thus greatly reducing the overall power efficiency. At low energy, the need to provide many cold–warm transitions to accommodate the large number of warm quadrupoles required because of the short focusing periods increases the cost and complexity of the installation. On top of that, the difficulty in predicting the individual gradients of superconducting cavities makes them less attractive at low energy, where as we have seen the sequence of cell lengths has to precisely follow the calculated increase in beam velocity.

The result is that while superconductivity is certainly the most attractive technology for linacs at high energy and high duty cycle, at low energy and low duty cycle normal-conducting structures remain more economical and more efficient. An exact comparison of cost and efficiency for the two technologies is particularly difficult because it depends not only on energy and duty cycle, but also on other design parameters such as repetition frequency, peak beam current and pulse length. A high repetition frequency makes a superconducting linac less efficient, more power being lost during the long pulse rise time required to feed the superconducting cavities. The maximum current during pulse plays an important role, because whereas normal-conducting linacs are more efficient operating with short pulses of high beam current, superconducting linacs prefer long pulses with less current that require a smaller RF installation. For all of these reasons, the optimum transition energy between warm and cold sections in a modern linac remains difficult to determine and requires a precise economical comparison of the two technologies for the parameters of each particular project. A special case are linacs operating in CW, where the tendency nowadays is to start the superconducting section immediately after the RFQ, usually at 3 MeV. Although the general rule remains that the higher the duty cycle the lower the optimum transition energy, different projects have chosen different transitions; as an example, Fig. 26 plots the selected transition energy as a function of duty cycle for the most important linacs built in recent years or in the design and construction phase. The approximate trend line connects the low transition energy of CW projects with the high transition energies of low duty cycle projects and represents the “state-of-the-art” in warm–cold transition.

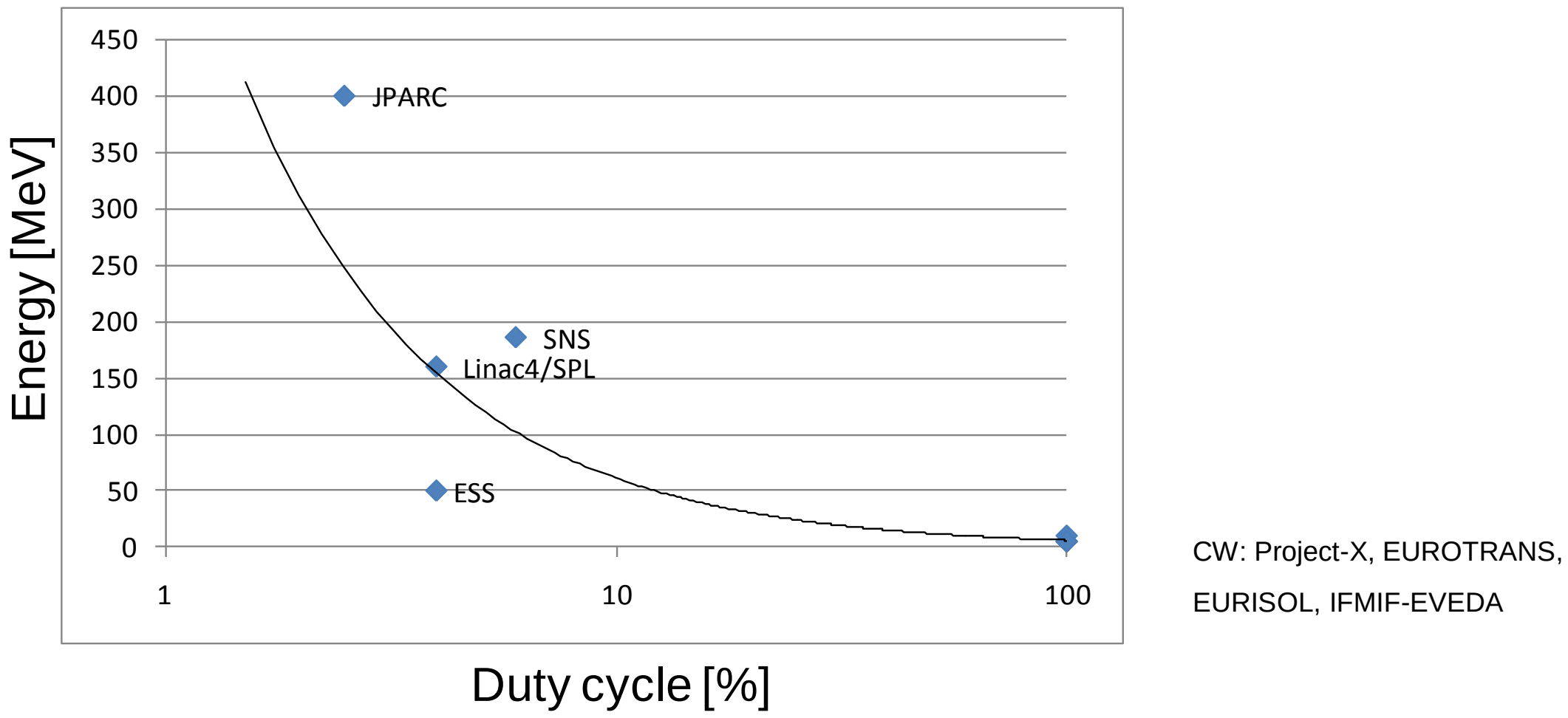

**Fig. 26:** Warm–cold transition for recent linac designs (built or in the construction or design phase)